\documentclass[structabstract]{aa}  
\usepackage{graphicx}
\usepackage{lscape}
\usepackage{longtable}
\usepackage{natbib}
\usepackage{color}
\usepackage{array} 
\bibpunct{(}{)}{;}{a}{}{,} 

\usepackage{txfonts}
\begin{document}
   \title{Multiplicity in transiting planet host stars}

   \subtitle{A Lucky Imaging study of Kepler candidates}

   \author{J. Lillo-Box
          \inst{1}
          , D. Barrado\inst{1,2}, H. Bouy\inst{1}
          }

   \institute{Departamento de Astrof\'isica, Centro de Astrobiolog\'ia, ESAC campus 28691 Villanueva de la Ca\~nada (Madrid), Spain\\
              \email{Jorge.Lillo@cab.inta-csic.es}
         \and
             Centro Astron\'omico Hispano-Alem\'an (CAHA). Calar Alto Observatory, c/ Jes\'us Durb\'an Rem\'on 2-2, 04004, Almer\'ia, Spain.\\
             }

   \date{Received 19 May 2012; Accepted 1 August 2012} 

 
  \abstract
   {In the exoplanetary era, the Kepler spacecraft is causing a revolution by discovering thousands of new planet candidates. However, a follow up program is needed in order to reject false candidates and to fully characterize the bona-fide exoplanets.}
   {  Our main aims are: 1./ Detect and analyze close companions inside the typical Kepler PSF to study if they are the responsible of the dim in the Kepler light curves, 2./ Study the change in the stellar and planetary parameters due to the presence of an unresolved object, 3./ Help to validate those Kepler Objects of Interest that do not present any object inside the Kepler PSF and 4./ Study the multiplicity rate in planet host candidates. Such a large sample of observed planet host candidates allows us to do statistics about the presence of close (visual or bounded) companions to the harboring star. }
   {We present here Lucky Imaging  observations for a total amount of 98 Kepler Objects of Interest. This technique is based on the acquisition of thousands of very short exposure time images. Then, a selection and combination of a small amount of the best quality frames provides a high resolution image with objects having a 0.1 arcsec PSF. We applied this technique to carry out observations in the Sloan i and Sloan z filters of our Kepler candidates.}
   {We find blended objects inside the Kepler PSF for a significant percentage of KOIs. On one hand, only 58.2 \% of the hosts do not present any object within 6 arcsec. On the other hand, we have found 19 companions closer than 3 arcsec in 17 KOIs. According to their magnitudes and $i-z$ color, 8 of them could be physically bounded to the host star.}
   {}


   \maketitle
%

\section{Introduction}

It is not so long when the discovery of extrasolar planets was just utopian. However, after the first discovery of an exoplanet orbiting a main sequence star \citep{mayor95,marcy96}, the scientific community has discovered and confirmed hundreds of these objects around other stars. In this context, the Kepler Space Telescope  has detected a new large sample of planet host candidates by continuously observing more than 150000 stars in a specific field of view (RA=19h 22m 40s DEC=+44 30' 00''). During the first five quarters of observations (i.e, $\approx 4.5$ months) the Kepler Team collected on its second public release a total amount of 997 planet host star candidates \citep[second public release,][]{borucki11}. 

However, to date, less than 5\% of these candidates have been confirmed. After the candidates selection, there is a mandatory step to reject false positives before attempting any high accurate (difficult and expensive) radial velocity measurements. Since the Kepler effective point-spread function is very large (6-10 arcsec, depending on the particular target) and its pixel size is about 4 arcsec, several background objects could be blended by the host candidate (called Kepler Object of Interest, hereafter KOI). Bounded or not, these objects clearly affects the star and planet parameters such as the planet-to-star radius ($R_p/R_*$), semi-major axis ($a/R_*$), impact parameter ($b$) or even the planetary mass ($M_p$). 

The presence of a secondary star could lead to the definite rejection of the candidate, as an example, \citep[see][]{odonovan06} In fact, there are several configurations that can mimic an exoplanet transit around its host star. The most relevant are: A./ a small substellar object transiting the other component of the binary system (since the smallest stars and brown dwarfs have the same size as Jupiter), B./ a stellar binary blended with a background star, C./ a grazing binary, which has not been ruled out by additional photometry or spectroscopy, D./ a background eclipsing binary blended by the light of the KOI, E./ a larger planet transiting a background star blended by the main target (this actually would not strictly be a false positive since there is a planet in the system but it would be in the sense of rejecting the brightest central star as a planet host) and  F./ a long-term spot. There would be a 'G-case' where the main target actually hosts a planet but with a blended background companion or a non transiting bounded companion. These configurations would lead to a change in the planet properties, as it was said before.  

Some of these configurations might be ruled out by the automatic pipeline implemented by the Kepler Team \citep{jenkins10}. While cases A and B are well rejected by this pipeline and an individual study of the light curves done by the team itself \citep{borucki11},  low-resolution spectroscopy clearly reject the C configuration. However,  D, E and G cases are the main sources of false positives in the sample of transiting planet candidates. More specifically, case G clearly shows the need for an intense high resolution imaging follow-up program to validate the planetary nature of the transients. Due to Kepler long base-line, we expect few or no case F.

Theoretical studies in regard to the false positive probability of Kepler candidates conclude that obtaining high resolution images down to 1-2 arcsec is crucial to confirm the planets and its physical properties. As an example, an earth-size planet transiting a faint star might have a false positive probability greater that 20\% if it lacks high-resolution imaging,  which could potentially be decreased to less than 2\% with a high-resolution image \citep{timothy11}. Several authors have acquired this kind of observations for other planet host candidates finding important changes in the planet-star properties. For instance, \cite{daemgen09} found stellar companions to 3 stars harboring planets. As a consequence, the updated values of the physical parameters changed about 2\% with respect to the previous ones.

However, even if the planet is confirmed, its formation and evolution scenarios (including the migration) require accurate description of the effect of bounded stellar companions. The vast majority of planets found in multiple systems are actually S-type, meaning that the planet is orbiting one of the components of the multiple system playing the secondary star the role of a gravitational perturber \citep{kley10}. The presence of these secondary objects difficult the planet formation since they interact dynamically with the elements of the system and produce an extra heating of the protoplanetary disk. All these factors may imply important changes in the planetary architecture and exoplanet properties with respect to those formed around single stars. For instance, \cite{eggenberg04} found a statistical segregation in the planet mass for those planets with orbital periods shorter than  40 days around single and multiple systems.

In this paper, we present the first statistical study of multiplicity on the Kepler candidates. A set of high resolution images obtained with the Lucky-Imaging technique \citep{law06} in the 2.2 m telescope in Calar Alto Observatory (Almer\'ia, Spain) with the AstraLux instrument were acquired.  This technique allows us to obtain diffraction limited observations with the best seeing conditions in the SDSSi band (see the F\'elix Hormuth PhD dissertation). A total amount of 98 KOIs (i.e., about a 10\% of the candidates listed in the Kepler second public release of the Kepler Team) have been pointed and studied.

In section \S2 we will explain the observations, image processing and data extraction from the raw images, regarding: sample selection (\S2.1), data acquisition (\S2.2), data reduction and photometric calibration (\S2.3), including the astrometric corrections, and a study of the sensitivity curves and detectability of our observations. Results based on these observations will be presented in section \S3. We also perform spectral typing of the detected companions (\S4.1) and give some clues about the possible gravitational bound between some of them (\S4.2). In section \S4.3 we will discuss the implications on calculated planetary parameters from the Kepler light curves regarding the presence of a blended star. Some particularly interesting cases will be studied in \S4.4 and final conclusions of this work will be presented in section \S5.


\section{Observations \& Data Reduction}

\subsection{Sample selection}

We have observed 98 KOIs. The majority of them were selected due to the lack of any kind of follow-up observations according to the published data of the Kepler Team \citep[, second public release]{borucki11}. The selection of the rest was coordinated with the Kepler Team. In addition, we limit our targets to be brighter than 15 magnitudes in the Sloan i band  in order to detect objects 5 magnitudes fainter than the KOI (see section \S~\ref{sensitivity}). Apart from these constrains, KOIs were randomly selected from the remaining sample so that we expect no bias in our results.


\subsection{Data acquisition}

The data presented in this paper was  taken in the Calar Alto Observatory (Almer\'ia, Spain) within 19 nights of observations divided in three separate runs (06-12 June 2011, 01-10 July 2011 and 25-26 July 2011). Although two nights were lost due to weather conditions, the remaining 17 had very good atmospheric stability with low atmospheric extinction and sporadic or absent clouds, ideal for this observing technique. The mean natural seeing was around 0.9 arcsec over the entire run.  

We used the AstraLux instrument placed at the 2.2 meter telescope to perform diffraction limited imaging of our sample of KOIs with the Lucky Imaging technique. We used the entire available field of view of the camera (i.e., 24 $\times$ 24 arcsec) to cover a separation range encompassing the entire mean Kepler PSF (6-10 arcsec).  In brief, we took thousands of images with short exposure times ($100-200$ milliseconds), well below the typical atmospheric turbulence changes (see Felix Hormuth PhD dissertation). We set the exposure times and number of frames according to the atmospheric conditions and target magnitude in order to reach at least 5.0 magnitudes fainter than the central KOI. Thus, our total exposure times are in the range 340-2000 seconds We note that fainter objects than this magnitude difference would not affect the planet-star properties by more than 0.5\% in the case of the $R_p/R_*$ rate, see equation (6) in \cite{seager03}. Moreover the probability that Kepler observations have detected a transit with a signal to noise ratio greater than 7.1 decreases as the star gets fainter. Thus, we design our observations to reach  $\Delta m_i \approx 5.0$  mag at 1.0-1.5 arcsec from the KOI.

Images were firstly obtained in the SDSSi filter since the PSF deformation is lower for this band than for shorter wavelengths (this result is quite similar for the SDSSz band). An on-the-fly reduction allow us to check inmediately for close sources. If any, SDSSz band images were acquired in order to characterize the secondary object by using the $i-z$ color (see section \ref{detection}).

\subsection{Data reduction and photometric calibration.}

		 \subsubsection{Basic reduction} We used the AstraLux pipeline (see AstraLux Diploma Thesis by Felix Hormuth\footnote{http://www.mpia-hd.mpg.de/ASTRALUX/}) to perform the basic reduction and combination of our lucky imaging frames. First, this pipeline applies the bias and flat field correction correspondent to each night to the science images. We used dome flat field to avoid inhomogeneities due to the pass of some clouds during the sunset in most of the observing nights. Second, it measures the quality of each science frame to select the 1.0\%, 2.5\%, 5.0\% and 10\% frames with the highest strehl ratios \citep{strehl1902}, calculates the shifts between those single frames, performs the stacking and divides the pixel size to half its value (i.e., the pixel scale is resampled from 0.0466''/pixel to 0.0233''/pixel). According to our observing configuration (number of frames, single exposure time and gain), we need to use the 10 \% selection rate images to reduce the photometric errors (as recommended in the User's Manual). Inspection of the lower selection rate images has been done for source detection purposes but not for photometric analysis presented in this paper. 
		 
		 \subsubsection{Astrometric corrections}
		 Astrometric corrections have been applied to the position of the targets when calculating angular separations and angles between them. We made use of our observations of the  M15 dense globular cluster. A total amount of 66 stars in our image were identified and matched with the \cite{yanny94} catalog from images of the Hubble Space Telescope. We performed astrometric calibration using the \textit{ccmap} package in IRAF. We compute the transformation between [x,y] coordinates to [$\xi$,$\eta$] tangent plane coordinates in a second order fitting: 

\begin{eqnarray}		 
		 \xi = c_{00}+c_{10}x+c_{01}y+c_{20}x^2+c_{02}y^2 \\
		 \eta = c'_{00}+c'_{10}x+c'_{01}y+c'_{20}x^2+c'_{02}y^2
\end{eqnarray}		 
		 
	where $c_{ij}$ are the correspondent calculated coefficients, which are shown in Table~\ref{astrometry}. We find residuals of $48.5$ mas in the $\xi$ direction and $34.5$ mas in the $\eta$ direction. Pixel scale and rotation angle are found to be $0.02359\pm0.0005$ arcsec/pixel and $1.78\pm0.01$ degrees, respectively. We will consider the rms value as our mean astrometric error in both tangential coordinates.

		\subsubsection{Source detection and photometric extraction} The SExtractor software for source extraction \citep{bertin96} was used to extract the pseudo-instrumental photometry (explained below) of sources surrounding each KOI. 
		We chose 60 pixels apertures (1.4 arcsec) to account for the maximum flux percentage of the objects. We set the detection threshold to 2.0-$\sigma$ in a minimum area of 30 pixels. Since this software does not take into account the exposure time and electromagnetic gain of the observations we had to correct the SExtractor output to obtain the real  instrumental magnitudes. The gain correction is necessary since different gains were set to different objects in order to avoid detector saturation. This was the case for standard stars for which we set the software gain in the range 230-255. For all the science images, a software gain of 255 was applied. However, this software gain is not the physical electron gain and it is not linearly correlated to it (see Figure 3.7 in the AstraLux Thesis). In a private communication, the PI of the instrument (Felix Hormuth) provided us with a calibrated function relating both parameters. Hence, calling Y to the physical electron gain, the instrumental magnitude of an object would be as follows: $m_{inst}=m_{SExtractor}+2.5log(Yt_{exp})$.
		
		
		 \subsubsection{Atmospheric extinction and Zero Point calculation} Extinction correction and determination of the zero points were done by using the observations of SDSS standard stars \citep{smith02}. We observed Ross 711 and SA105-815 during the first run; Ross 711, SA105-815 and BD+25-4655 in the second run; and Ross 711, BD+25-4655 in the third run. A good sampling of the airmass for each night ensured a correct determination of the calibration parameters for the photometric nights, namely the extinction coefficient ($\chi$) and the zero point ($C$). According to the Calar Alto weather monitoring webpage\footnote{publicly available at www.caha.es/WDXI/wdxi.php} and the results of the calibration diagrams ($m_{standard}-m_{inst}$ vs. {Airmass}) for each night, we conclude that the following nights were clearly photometric: July 5th, 6th, 7th, 10th and 26th. The average error in the calculated magnitudes for these nights is around 1\%. Another set of nights were classified as partially photometric because of sporadic clouds along the observing time or irregular seeing conditions that lead to an inaccurate photometry, since we observed the standard stars frequently. These nights were: June 11th and 12th, and July 1st, 3rd and 9th. Consequently, errors in these partially photometric nights are greater (1-3\%) but still acceptable for our purposes. The column \#5 in Table~\ref{results} provide information about the stability of the atmosphere for the observations of each object (according to the acquisition date). From our photometric results we can conclude that magnitudes brighter $i=21-22$ have assumable errors while fainter objects are too noisy to accurately determine their magnitudes with errors below 1.5\% and are then not taken into account for this study.
		 
		 Hence, the calibrated magnitude for these objects has been calculated as follows:
		 
\begin{equation}
m_{cal}=m_{inst}+2.5 log(A_p  Y)+Z_p+C\chi
 \end{equation}
 
 where $A_p$ is the aperture correction, Y  the electron gain, C the extinction coefficient,  $Z_p$ the zero point and   $\chi$ is the correspondent airmass. \\
		
When compared to the Kepler Input Catalog (KIC) values provided in the Kepler MAST archive\footnote{http://archive.stsci.edu/kepler/}, we obtain magnitude differences  ($i_{KIC}-i_{AstraLux}$) smaller than 0.1 mag for around 60\% of our objects without close companions. The 90\% of them are smaller than 0.4 mag.

		\subsubsection{Detection and Calibration of very close objects\label{detection}} 
		
		By looking into the reduced images, we selected the sub-sample of our KOIs that had close companions within 3 arcsec (17 KOIs) to perform an individual extraction of the photometry. Due to its proximity to the main target, we would need smaller apertures in order to not being contaminated from the PSF wings of the KOI. Hence, aperture correction was applied. Contamination coming from the KOI's PSF wing has been estimated by measuring the amount of flux from the KOI inside the companion's aperture. This extra flux would influence the magnitude of the companion in less than 0.03 magnitudes which is well below the photometric errors of these sources. Hence, no corrections of this nature have been performed in this work.
		Photometric extraction of the standard stars was also done with the IRAF package and calibration parameters were re-derived and applied to the science photometry. We determined the aperture correction by using the standard stars. As the \textit{phot} task of the  IRAF package already takes into account the exposure time of the image, here we just have to correct for: aperture correction, electron gain and atmospheric extinction. Hence, the final expression to derive the calibrated magnitudes is: 
		
\begin{equation}
m_{cal}=m_{IRAF}+2.5 log(A_p  Y)+Z_p+C\chi
 \end{equation}
 
 where $\chi$ is the correspondent airmass. 

{We also tested a PSF approach rather that aperture photometry. The main problem of this technique with Lucky-Imaging observations is to find a standard PSF for the whole night. Due to the acquisition technique, each particular image/star would have a characteristic PSF that cannot not be modeled by a general one. However, aperture photometry for very close ($<$1.5 arcsec) and faint companions can involve larger errors even using small apertures. Hence, we decided to use PSF photometry for the 2 KOIs with faint companions  at very small angular separations (KOI-1375 and KOI-0387). The PSF was obtained from different available standard stars during the whole night. We calculated the stellar fluxes for each PSF and estimate the relative magnitudes between the KOI and the companion ($\Delta i$ and $\Delta z$) as a weighted average of all values. Since aperture and PSF photometry must coincide for the bright KOI, we used the calibrated photometry for the KOI obtained by aperture photometry ($m_P$) to obtain the companions' magnitudes as $m_{/C}=m_P+\Delta m$ for each filter in these two cases. For the remaining objects, we used aperture photometry.\\

\subsubsection{Sensitivity, detectability and limiting magnitudes \label{sensitivity}}
 
 We have estimated the mean completeness and limiting magnitudes of our images by using the observations of the globular cluster  M15 in order to have a large sample of stars with a wide range of magnitudes. We measured the number of objects per magnitude bin of 0.5 mag on every photometric night for which images of M15 were taken. Computing the mean values of these bins we can construct the histogram shown in Fig.~\ref{depth}. We have scaled this histogram to a 200 seconds exposure time image to account for the mean exposure time of our science images. We obtain a mean completeness value of $i_{complete}=18.4\pm0.3$ mag and reach detectability down to approximately $i=22.5$ mag. In the case of the SDSSz observations we adapted the exposure time to detect the companion seen on the SDSSi image. It is important to remember that these values have been calculated for a particular exposure time. Since we set different exposure times for each target in order to achieve $\Delta m_i\approx 5.0$ (see lower panel in Fig.~\ref{figureresults}), completeness and detectability limits should change for each image. The scaled values of $i_{complete}$ for each particular science image attending to its real exposure time are shown in Table~\ref{obsresumen}. Note that we only show here the observations for the non-isolated KOIs (Kepler host candidates without any objects closer than 6 arcsec).
\addtocounter{table}{1}

    \begin{figure}[Ht]
   \centering
   \includegraphics[width=0.5\textwidth]{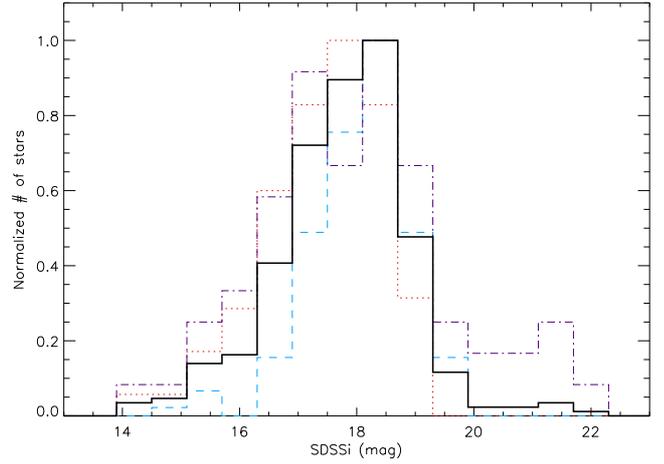}
      \caption{Mean completeness and detectability limits of our survey (black solid line) for a 200s effective exposure time image. Red dotted line, blue dashed line and purple dotted-dashed line represent the values for the 3 photometric  nights considered, being: 26jul11, 05jul11 and 07jul11, respectively. Bin size is 0.5 magnitudes. The histograms have been normalized to their maximum value for visualization purpose.}
         \label{depth}
   \end{figure}

Sensitivity functions were compiled for each image to determine the depth in angular separation and magnitude (actually, $\Delta m$) to which we are complete for $3-\sigma$ detection threshold. 

By measuring the observed radial profile averaged over a large number of directions (avoiding those angles in the way to the close companions) we can infer the dependence of the primary star brightness along the angular separation. Then, we reiteratively add the same profile but located at different angular separations between 0.2-3.0 arcsec and scale it to be $\Delta m$ magnitudes fainter according to the expression: $F_{/C}=F_P 10^{-0.4\Delta m}$, being $F_{/C}$ the encircled flux of the companion and $F_P$ the flux of the primary.  We then measure the signal to noise ratio (SNR) for the detection of the added profile. Figure~\ref{sens} shows an example of these calculations where the  artificial companion to the observed KOI profile range differential magnitudes of $\Delta m=[0,7]$ and angular separations between 0.2-3 arcsec. 


    \begin{figure}
   \centering
   \includegraphics[width=0.5\textwidth]{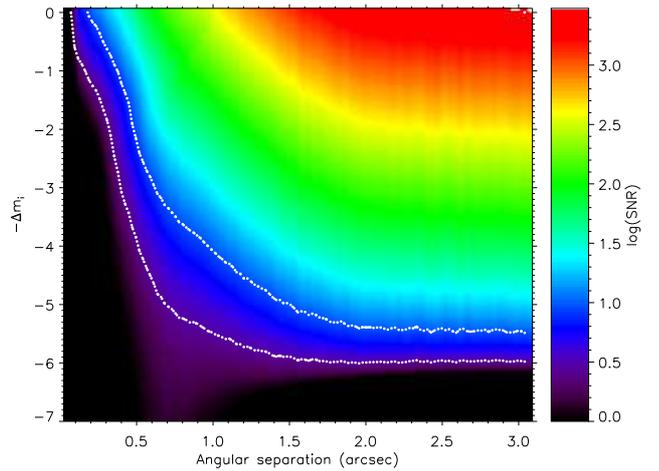}
      \caption{An example of the sensitivity function for our AstraLux observations. We show the results for KOI-0717 for an image with the 10\% selection rate. Color code represents the measured signal to noise ratio for the detection of the artificial companion profile. The white line represents the contour for the $3-\sigma$ (lower line) and $10-\sigma$ (upper line) SNR detection.}
         \label{sens}
   \end{figure}



\section{Results}

We have found 111 companion sources at angular separations between 0.3-10 arcsec from the 98 targeted KOIs. 
Since the Kepler point spread function varies from 6 to 10 arcseconds, we consider a KOI to be isolated if no objects  below 6 arcseconds are found in our Lucky Imaging observations. Objects with larger angular separations could be easily detected by the Kepler images or previous ground-based all sky surveys. 

According to this definition, we have detected 57 isolated KOIs. This means  a 58.2\% rate of isolated objects for planet host candidates in the Kepler Objects of Interest catalog. It is important to remark that this isolation rate would decrease down to the 33.7\% if we consider a 10 arcsec PSF for Kepler observations. However, the best PSF assumption of 6 arcsec will be considered in this paper. 


Regarding KOIs with visible companions inside 6 arcsec, 41 planet host candidates have, at least, one object within this projected separation (i.e., $41.8$ \%). Among them, 17 KOIs present a total amount of 19 companions at less than 3 arcsec (17.3\% of all KOIs considered). We will refer to this group along this paper as the close companions' group. In Fig.~\ref{close17} we show the 17 KOIs with close companions and the Kepler optimal (public) apertures of the different Quarters in which the Kepler Mission is divided. All 19 companions lie inside these apertures contaminating the light curve fluxes with its relevant impact in the planet-star properties and planet validation. Moreover,  27 KOIs do present 30 companions within the range 3-6 arcsec  (27.6\% of all KOIs considered), medium distance group. This means that 3 of the main targets from the close companion group ( KOI-0433, KOI-0641 and KOI-0841, see Fig~\ref{comp036}) present either close ($<$ 3 arcsec) and medium-separation  (3-6 arcsec) sources. Figure~\ref{piechart} and Table~\ref{mainresults} summarize these results.

According to all these numbers, we have defined the observed companion fraction as $ocf=\frac{n}{s+n}$, where n is the number of objects within a certain angular separation to the KOI for each particular type of system (either double, triple or quadruple) and $s$ is the number of isolated KOIs found in our sample (i.e., $s=57$). The cumulative $ocf$ is plotted in the upper panel of Fig.~\ref{figureresults}. We show there that given a KOI, the probability of having a double (visual) system within 3  arcsecs is 21\% while if we go further away until 6 arcsecs, this probability increases to 37\%. The same was done for triple systems (purple in upper panel of Fig.~\ref{figureresults}) and we found that only the 3\% of the KOIs do present more than 1 visual companion within 3 arcsec. No quadruple systems were detected in our sample within 6 arcsecs (orange line in the mentioned figure). We have also plotted a total observed companion fraction (black line) taking into account contributions from all type of systems. It is defined as $ocf(Total)= \frac{d+t+q}{s+d+t+q}$ where d, t and q are the double, triple and quadruple systems found within a certain angular separation \citep[similar definition as the one presented by ][]{duchene99}. Note that in the upper panel of Fig.~\ref{figureresults} only the $ocf_{Total}$ is monotonically increasing as we move away from the primary. The decreasing cumulative $ocf$ at 7-8 arcsec for the Double systems is due to the fact that when we reach the position of the third component of a triple system, this is automatically accounted in the Triple sample and removed from the Double sample. The same reason explains the decreasing cumulative $ocf$ in the Triples system. 

\begin{figure}[ht]
\centering
\includegraphics[width=0.5\textwidth]{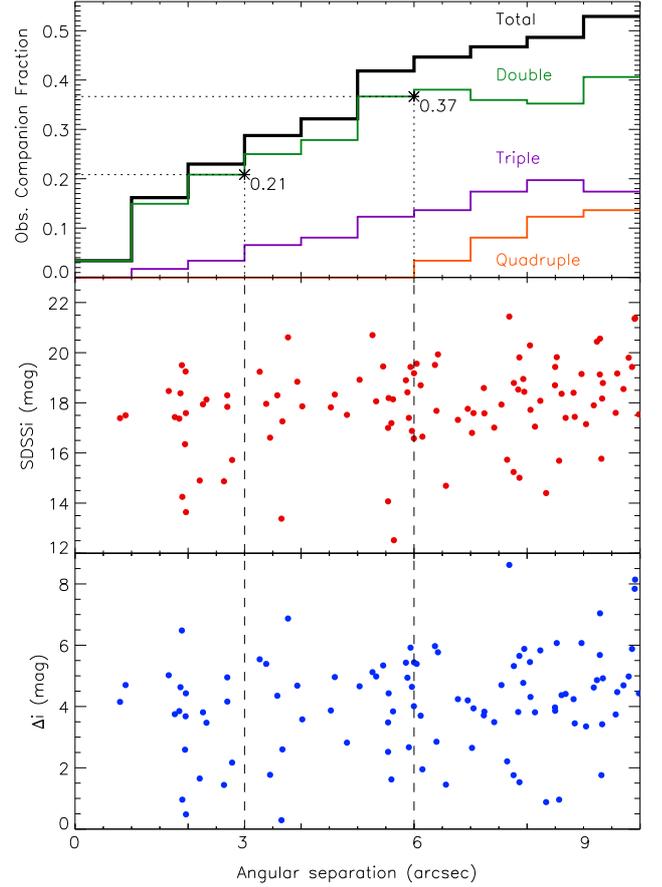}
\caption{\textbf{Upper panel:} Cumulative observed companion fraction ($ocf$, see text) for the double (green), triple (purple) and quadruple (orange) systems as well as the total $ocf$ (black). We also show the values for the double system at 3 and 6 arcsec. \textbf{Middle and lower panels:} calibrated SDSSi magnitude of the 111 companions and differential magnitude with their KOIs. Dashed lines mark the 3 arcsec and 6 arcsec positions. }
\label{figureresults}
\end{figure}

\begin{figure*}[Ht]
\centering
\includegraphics[width=1\textwidth]{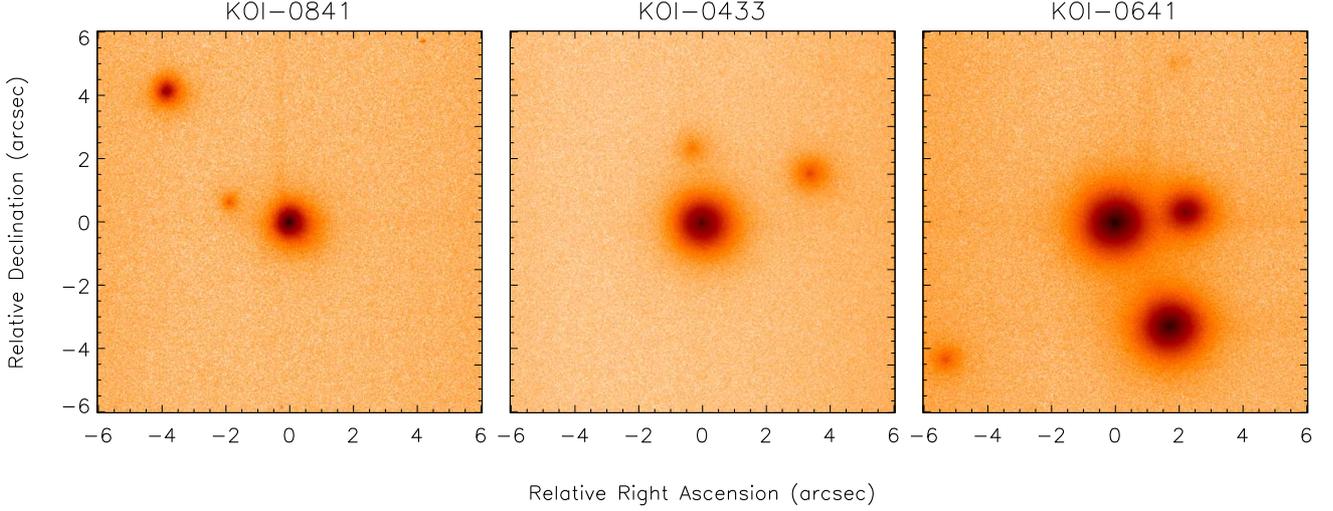}
\caption{High resolution images of the three KOIs with companions as in the 0-3 arcsec range as in the 3-6 arcsec one. The images are $12\times 12$ arcsec  and the KOI is placed at the center of the image. Companions are sorted by angular separation in Table~\ref{results36}.  North is up and East is left.}
\label{comp036}
\end{figure*}

\begin{figure*}[ht]
\centering
\includegraphics[width=17cm]{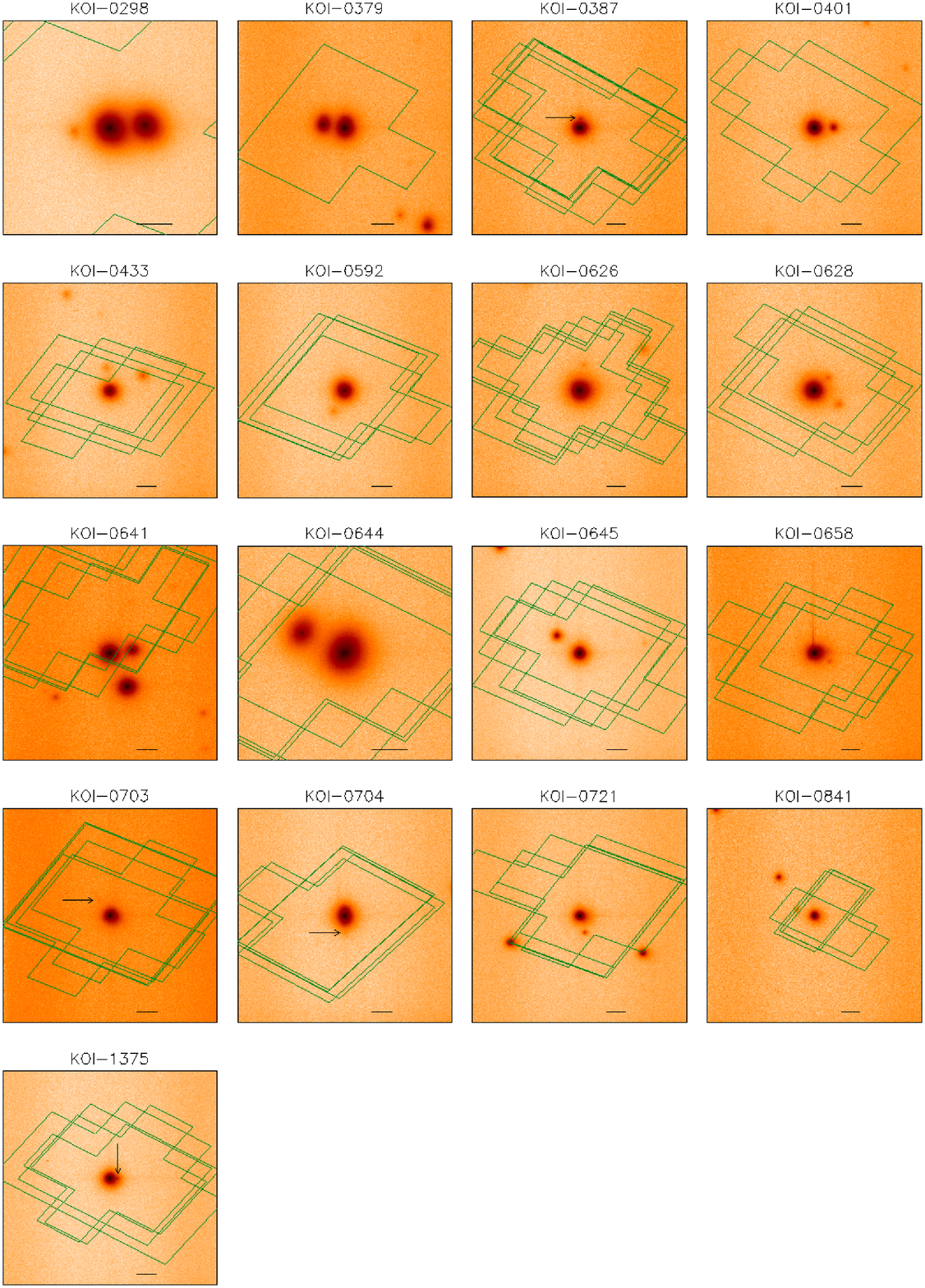}
\caption{High resolution SDSSi images of the close companions detected around 3 arcsec of 17 Kepler Objects of Interest. North towards up and East towards left.  Different sizes are used for each image to clearly show all features. Black arrows indicate the position of the faintest targets and  the horizontal black line represents 2 arcsec on each image.  Green polygons represent the optimal public Kepler apertures used on each Quarter of the Kepler Mission to compute the light curves.The brighter object assumed to be the KOI is centered in the images for reconnaissance purposes.}
\label{close17}
\end{figure*}


In Table~\ref{results}, \addtocounter{table}{1}  we present our photometric measurements in the SDSSi and SDSSz bands for the KOIs with detected close ($< 3$ arcsec) companions. In this table we also provide the angular separations between the stellar companions and the assumed KOI as well as  the angular position from North towards East. It is important to notice that in cases where both objects have similar magnitudes it is not possible with our observations to infer which of the two is actually the one that harbors the transit (high resolution imaging in- and out-of-transit will be performed to that end).  In order to avoid errors coming from the calibration process, we have measured the differential magnitudes $\Delta i$ and $\Delta z$ directly from the instrumental magnitudes. The $i-z$ color has been derived directly from the calibrated magnitudes for the primary (brighter) star. Instead,  in order to minimize the large errors due to the faintness of the majority of the close companions, we obtained the $(i-z)_{/C}$ color of the companion by using the expression:

\begin{figure}[ht]
\centering
\includegraphics[width=0.5\textwidth]{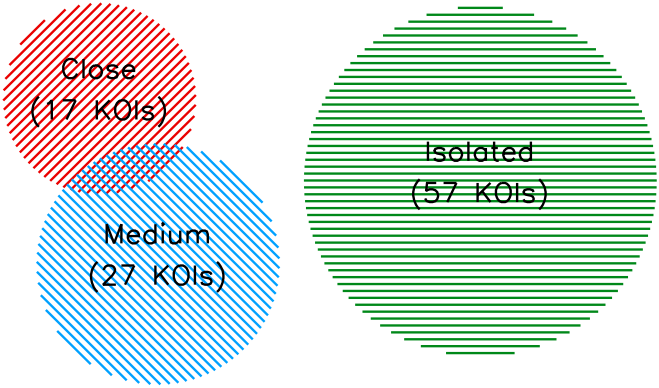}
\caption{ Main multiplicity results of our survey. Isolated KOIs are represented by the green color. KOIs with objects between 3-6 arcsec are represented by the blue color. The close companion  group (KOIs with at least one source between 0.3-3.0 arcsec) is shown in red. Note that 3 KOIs (KOI-0433, KOI-0641 and KOI-0841) have objects either in the close and medium distance groups.}
\label{piechart}
\end{figure}

\begin{equation}
(i-z)_{/C}=\Delta i - \Delta z + (i-z)_P 
\end{equation}  

where $(i-z)_P $ represents the color of the primary star. 

Due to effectivity reasons in our limited observing time, we have only obtained SDSSz images for those KOIs showing very close companions ($< 3$ arcsec) in an on-the-fly reduction. Hence we lack $i-z$ colors for companions at 3-6 arcsec. Photometric and astrometric information available from our observations for these medium-separation sources is shown in Table~\ref{results36}\addtocounter{table}{1}. Note that KOIs with companions at 3-6 arcsec that also have at least one object in the 0.3-3 arcsec,  do have SDSSz photometry.   

Reconnaissance spectra has been taken for  25 KOIs in order to more accurately derive their spectral types and physical parameters as temperature and surface gravity. Moreover, as said in \cite{borucki11} this kind of observations are able to reject the possibility of  a very close ($< 0.2$ arcsec)  binary that would produce a grazing  eclipse, since radial velocities of tens of km/s would be present on it. The data is now being analyzed and will be published in a future paper (Lillo-Box et al., 2012b, in prep.).\\

%


\section{Discussion}


\subsection{Spectral types of the stellar components}

We have applied two different methods to determine the spectral type (effective temperature) of the stellar companions according to their angular separation.  

\subsubsection{Spectral Energy Distributions for 3-6 arcsec companions.}
Among the 27 KOIs with stellar companions at 3-6 arcsec, we find 2MASS counterparts \citep{cutri03} for around one third of the detected companions (11 out of 30). We have assumed an error of 1.5 arcsec for the cross-match between catalogs.  Visual inspection of the 2MASS images have been carried out to ensure the accuracy of the match. Together with our SDSSi magnitudes we have performed a spectral energy distribution (SED) fitting with the Virtual Observatory SED Analyzer tool  \citep[VOSA,][2012 in prep.]{bayo08} in order to determine the effective temperature of our objects. The new version of this tool allows us to also fit the extinction along the line of sight. The results show values for the $A_v$ parameter in the range $A_v=0-1.0$. Here we have assumed solar metallicity and two surface gravities  to account for two possible evolutionary stages of the companion: Main Sequence foreground or bounded stars ($log(g)=4.5$) or background giant star ($log(g)=3.5$). Table~\ref{VOSAresults}  \addtocounter{table}{1}  summarizes the fit results. According to them, these companions are mostly K-type stars if we assume a main sequence stage. Note that all KOIs are hotter (earlier spectral types) than the possible companions. We find an $rms\sim200 K$ between our fitting results and the KIC effective temperatures. Constrains on the distances for these objects are explained in section \S~\ref{BoundedMedium}

\subsubsection{Spectral types for close companions.}
Assuming a Main Sequence nature for the secondary,  we can estimate their spectral types by using the $i-z$ color. Synthetic spectra from \cite{pickles98} for main sequence stars were convolved with the SDSSi and SDSSz transmission curves\footnote{see http://www.sdss.org/dr3/instruments/imager/index.html}  following the same scheme as in \cite{daemgen09} to derive a relation between the spectral type and the mentioned color.  We can over plot our $i-z$ colors in this diagram to estimate the companion spectral types. Typical errors depend on the spectral type since they are determined according to the photometric error in the calculus of the $i-z$ values. In particular, it is important to notice that, since the dependence of the spectral type with the mentioned color is more steeped for types later than K5-K6, errors will be much smaller than for earlier types where the dependence starts to flatten. Thus, only stellar companions with $i-z>0.21$ (correspondent to a K5-K6 main sequence star) are considered in this analysis. In Fig.~\ref{spt} we show the results for the 9 companions and 2 primaries with estimated spectral types later than K5 (see Table~\ref{companionspt}).

As expected, these objects are cool, low-mass stars (late K or early M). Their physical association will be discussed in the following section.

    \begin{figure}
   \centering
   \includegraphics[width=0.5\textwidth]{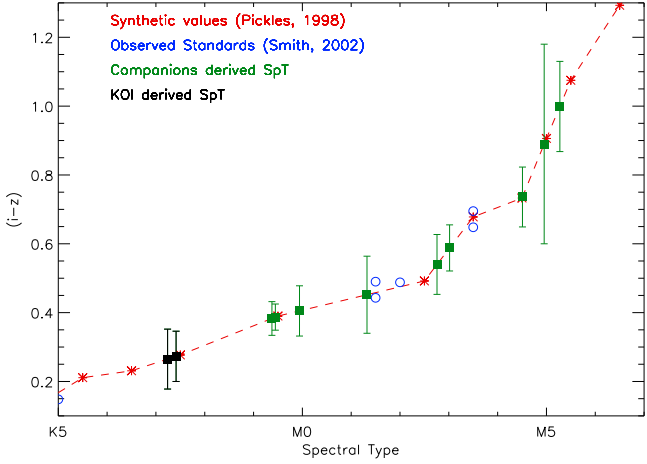}
      \caption{Spectral type estimation for the stellar companions with  $i-z>0.21$ corresponding to spectral types later than K5-K6.  Red asterisks represent the synthetic values calculated by convolving the spectral library from \cite{pickles98} with the transmission curves of the Sloan filters. Blue open circles represent the position of standard stars  from \cite{smith02}. Green filled squares are the measurement for our detected companions and their error bars (assuming no error for the software gain).} 
         \label{spt}
   \end{figure}

\subsection{On the physical association of the visual companions} \label{boundedornot}
Although the physical bound of the blended stars do not affect the previous calculations, it is important to determine if the visual companions are, indeed, bounded or not. Recently, several multiple-star systems have been discovered, both circumbinary planets as, for instance, Kepler-16b \citep{doyle11} and  binary systems with the planet orbiting one of the components of the couple as HD196885Ab \citep[see][]{thebault11}. These objects represent a challenge for theoretical models of planet formation, for instance, in terms of orbital stability \citep[see]{holman99}. Moreover, it has been discovered that around one fifth of the known exoplanets inhabit double or multiple stellar systems \citep{desidera07,mugrauer09,thebault11}.

\subsubsection{Medium-distance companions \label{BoundedMedium}}
We can infer rough distances to the KOIs and companions analyzed in previous section by a SED fitting by using the total flux emitted by the star ($F_{tot}$) according to the effective temperature estimated by VOSA. From this temperature and assuming a luminosity class we can estimate the bolometric magnitude of the star by using the relations in \cite{schmidt-kaler82}. Hence, a luminosity $L_*$ can be derived so that we can evaluate the distance since $L_*=4\pi d^2F_{tot}$. For the KOIs we have assumed a main sequence stage but two distances have been calculated for the stellar companions assuming main sequence and giant stages. The errors have been calculated by taking into account the uncertainties in $F_{tot}$ and assuming 125 K of error in the temperature calculation for the $L_*$ derivation. The former uncertainty also includes the errors in the effective temperature and  metallicity fitted by VOSA. The results are shown in the last column of  Table~\ref{VOSAresults}. 

The giant scenario for the companion does not seem to work with most companions since very large distances are found ($d>21Kpc$, the Galaxy limit in the Kepler line of sight). Thus, the main sequence scenario is the only possibility for them. According to it, one (KOI-0623C) out of the 10 companions that could be fitted by VOSA has a distance in good agreement with its correspondent KOI.  Among the remaining 8 possible companions, 7 are found to be background main sequence stars and one (KOI-212C) is a probable foreground M-dwarf with spectral type M4V.

\subsubsection{Close companions \label{closebounded}}

\cite{timothy11} estimated that the probability for a given target in the Kepler field to have a background source brighter than  $m_k=24.0$ within 2 arcseconds. This value strongly depends on the kepler magnitude ($m_k$) and galactic latitude ($b$) of the target. Since the 90\% of our primary targets have kepler magnitudes in the range $13.0-15.5$ and taking into account equation [8] in the mentioned paper, we can derive a background source probability range of 4-40\% depending on the particular values of $b$ and $m_k$ for each target. However, as it was previously said, we reach completeness for $i=18.5$ (which translates to $m_k=18.8$  according to a simple transformation with $r^2=0.97$ using both KIC magnitudes) so that we would underestimate the chance-aligned probability with respect to the one in the mentioned paper above. We have found 11 blended sources at less than 2 arcsec in 10 parent stars. That means a 10.2\% of our sample which, even being a lower limit, clearly agrees with the mentioned values estimated by  \cite{timothy11}.


Even though our photometric measurements do not provide enough information to clearly determine the gravitational bound (if any) of the system, we can infer some hints of this fact by analyzing several aspects of the available magnitudes of the A and B/C components. We have constructed an empirical Zero Age Main Sequence (ZAMS) based on the synthetic $griz$ photometry derived by \cite{ofek08} for Tycho-2 stars with Hipparcos $B_T$ and $V_T$ bands as well as 2MASS JHK  magnitudes (see Fig.~\ref{zams}). In order to reach the substellar  domain, we also add the \cite{moraux03} $iz$ photometry for 109 brown dwarfs in the Pleiades region assuming a distance to the cluster of 130 pc \citep{stello01}. As we are interested in possible binary stars, we have computed the lower-envelope of the ZAMS for both catalogs together in an $i$ vs. $i-z$ color-magnitude diagram. Primary objects were located in that empirical ZAMS according to their $i-z$ color, calculated from our own photometry. Hence, a distance modulus can be calculated for the primary star and applied to the secondary star assuming a joint formation for the system. Figure~\ref{zams} shows the results for all of our 17 KOIs with their 19 close stellar companions. At least 6 companions agree with a simultaneous formation in a double system (KOIs 0379B, 0645B, 0298B, 0641B, 0658B and 0401B), together with the KOI star. Other 3 close objects could also  lie inside the ZAMS boundaries if we take into account the errors in the determination of this isochrone (KOIs 0592B, 0433B and 0703B). It is also  important to note that some of them lie in the low-mass domain of the diagram. According to it, KOI-0641B, KOI-0658B, KOI-0433B and KOI-0703B  could be low-mass stars or even brown dwarf companions to the host stars.

 \cite{brandner00} provided with a simple formula to estimate the probability for a given source at a certain angular separation to be a background object. If we assume the limiting magnitude of the  USNO-B1.0 catalog \citep{monet03} in the Kepler field to be $I=17.5$ magnitudes , we find an overall probability of 3.9\% for a source separated 3 arcsec from our target to be a background object. This shows, although as a first approximation, the low probability for our close companions to be background sources rather than bounded binaries. 

 Our analysis implies that 6-9 found companions out of the 98 KOIs observed could be physically bounded. This means an observational lower limit of 6.2\%-9.2\% for the binarity rate among the Kepler targets (binaries with projected separations  and/or magnitudes below our detection limits may increase these values). However, we warn this result regarding the lack of confirmation of their physical bound. Instead, we have shown several hints that point to this scenario. As it was said before, it has been estimated that 20\% of the planets inhabit binary systems, most of them being wide binaries with separations greater than 1000 AU \citep{thebault11,desidera07,mugrauer09}. This could be in good agreement with our results if the physical bound is confirmed for the 6-9 KOIs mentioned above. The presence of such widely separated companions would had weakly affected the planet formation. \cite{desidera07} concluded that companions at more than 100-300 AU would not significantly influence the formation and migration of planets and thus, planet parameters may not have relevant differences with respect to those inhabiting single stars except for an over-abundance of high-eccentricity orbits in the wide binary planets. We do not detect any differences when comparing the KIC derived physical and orbital parameters of the 6-9 KOIs to the rest of the observed sample, confirming the results previously mentioned. 
 
 We note that our estimated binary fraction represents a lower limit  due to our observational restrictions. Close and/or faint (low-mass) companions beyond our detectable limits (described in section \S 2.3.6) might increase this multiplicity fraction. \cite{raghavan10} obtained a binary frequency of 34\% for companions around close ($< 35pc$) solar-like stars in the magnitude range $V=4-12$. Although our present survey encompasses a different range of stellar masses, their result suggests that our study might indeed be missing a number of close and faint companions.

    \begin{figure}
   \centering
   \includegraphics[width=0.5\textwidth]{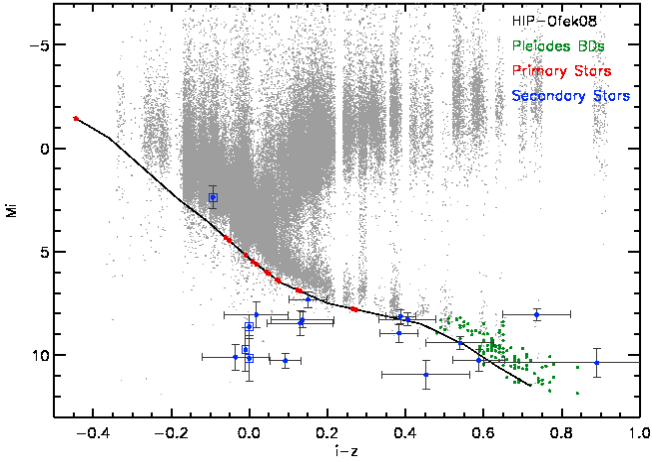}
      \caption{Empirical Zero Age Main Sequence (solid black line) computed by using synthetic  $iz$ photometry from \cite{ofek08}, grey dots, and \cite{moraux03}, green dots. Primary objects are represented by red filled circles and secondary companions with blue filled circles. Companions with large errors in the $i-z$ color value are plotted with a blue open square. }
         \label{zams}
   \end{figure}

\subsection{Need for update of planet-star parameters}

	The incidence in the planet and host star parameters due to the presence of a bounded or visual companion to the host star can be measured by isolating the contribution of the host star from the blended companion. Although the transit depth defined as $F_{notransit}-F_{transit}$ do not depends on the presence of blended objects in the PSF, physical properties of the planet-star system actually depend on the quotient $\Delta F=(F_{notransit}-F_{transit})/F_{notransit}$ as shown by \cite{seager03}. This quantity is clearly decreased with the closeness of a blended object since it would contribute with an additional flat flux (an intrinsic variability of the blended star would be visible in the light curve). Hence, a change of $(F_{notransit}-F_{blend})/F_{notransit}$ will imply consequent changes  in the planet-star parameters. Assuming that the planet is actually transiting the brighter star, the new depth will be given by \cite[see][]{daemgen09}:

\begin{equation}
\label{eqflux} 
\Delta F_{new}=\left( 1+10^{-\frac{\Delta z}{2.5}} \right) \Delta F_{old}
\end{equation}

It must be noticed that for the two systems where more than one blended object is found, the $\Delta z$ should be computed as: 

\begin{equation} 
z_{/C}'= -2.5 log\left[\sum_j{10^{-0.4z_j}}\right]  
\end{equation}

Additional photometric studies like on-transit and out-of-transit high resolution images of the fainter star as well as  observations of the transit depth at longer wavelengths such as with the IRAC bands of the Spitzer Space Telescope \citep[see, for example, ][]{desert11}  would be needed to rule out the possibility of the planet transiting the fainter star. 

The impact  on the physical parameters due to the presence of a blended companion could be quite significant. The largest change in the transit depth values according to Eq.~\ref{eqflux} is $\Delta F_{new} \le  \sqrt{2} \Delta F_{old}$, assuming an equal-magnitude blended star. According to equation (6) in \cite{seager03}, this would increase the planet to star radius ratio in a 41\%. The same calculation can be done for other parameters like:  total on-transit duration ($t_T$), total on-transit flat duration ($t_F$) or period ($P$). We have performed this exercise for our observations. Table~\ref{newparams} shows the re-calculated values of the transit depth, planet-to-star radius ratio  and planet radius for the 24 planets orbiting the 17 stars with close companions. The last two parameters have been calculated by assuming no-limb darkening, using the mentioned formula in \cite{seager03}  which reads: $R_p/R_*=\sqrt{\Delta F}$. Our results show relative differences in the transit depths, calculated as $(\delta_{new}-\delta_{cat})/\delta_{cat}$, in the range 1\%-120\% with respect to the values published by \cite{batalha12}. According to the expression explained above, the 74\% of the planets with detected companions at less than 3 arcsec would have changes in their $R_p/R_s$ parameter below 10\%, and the 91\% of the them below 30\%.}

\subsection{Particular cases \label{ParticularCases}}

\subsubsection{KOI-0641}

This system is of particular interests. It is a triple visual system  within 3.7 arcsec. The $i-z$ color study reveals spectral types of K5-K9, M1-M3 and M3-M5 for the A, B and C components respectively. Regarding the SED fitting, we obtain spectral types M2 and M3-M4 for the A and C components respectively. Distances to the primary target agree in both approaches: $d=125\pm 22$pc and $d=117^{+6}_{-5}$pc. The KOI-0641B perfectly suits the bounded scenario according to its position in the ZAMS. If the case, it would have a projected distance of $275\pm49$ AU to the A component. The third object is located well above the empirical ZAMS ($\sim 3$ mag. brighter), indicating that it probably is a foreground M-dwarf. This hypothesis is in agreement with the distance calculated by the SED fitting approach, $d=76^{+15}_{-10}$, significantly closer than the A component.


\subsubsection{KOI-0433}

The companion to KOI-0433 is too faint and red to extract conclusions about its nature. Our analysis suggests that it is a $M5^{+1}_{-2}$ star probably bounded to the main KOI. However, the primary target is also too faint ($m_i>14.5$) and large errors in the determination of the distance do not allow us to set constraints in the projected separation of the system. Even that, due to the position of the secondary object in the HR diagram of Fig.~\ref{zams}, as in KOI-0641C, there exists the possibility of having a third component very close to KOI-0433B not resolved in our images. Hence, we consider this object as a possible false positive although more observations need to be done. 

\subsubsection{KOI-0298}

The projected distance estimated for the assumed primary target is $118 \pm 18$ pc according to section \S~\ref{zams}. The stellar companion to this KOI has a $i-z$ color typical of early M dwarfs. Its position on the $M_i$ vs. $i-z$ diagram is in good agreement with a simultaneous formation of both objects. Thus, a relative separation of $231\pm 35 $ AU is derived for the possible binary system. Moreover, since both objects have very similar magnitudes, it is not clear which star is actually hosting the transient object. Note also the presence of a third very faint object in the SDSSi image. This object is not detected in the SDSSz image acquired  few hours later in the same night. The most probable reason is a background blue object.

\subsubsection{KOI-0379}

Its projected distance has been estimated to be $244 \pm 76$ pc. The posible companion with an spectral type earlier than K5, if bounded, would be located at  $463\pm 165$ AU from the primary target. However, since this estimations have been based of photometric values and the correspondent night was not photometric, we must warn that errors are probably larger than the ones provided here.  

\subsubsection{KOI-0645}

KOI-0645 is a two-planet candidate system. The location of the secondary star in the HR diagram suggests a physical bound with the candidate host star. Estimated distance is $328\pm 82$ pc leading to a relative separation of $910\pm 228$ AU.

\subsubsection{KOI-0401}

This is another star presumably hosting two planets.  The stellar companion has been estimated to be a K9 star at a projected distance of $590\pm 190$ AU if we assumed as valid the distance of $302\pm 98$ pc derived for the primary. 

\subsubsection{KOI-0623}

The nature of the stellar companion at 5.54 arcsec to the KOI-0623 still remains unclear in this work. Both assumed luminosity classes are suitable. If it was a giant K2 background star it should be at $790^{+250}_{-350}$ pc. Instead, the dwarf nature would allow a possible bound with the A component since their distances are very close although they lay outside the error boundary.

\subsubsection{KOI-0658}

A very faint companion with an approximated spectral type of M3 has been detected close to this KOI. Due to its faintness, large errors are present in the determination of the projected distance, being $790\pm330$ AU if we assumed a $420\pm180$ pc distance to the primary. Moreover, with a $\Delta m_z\approx 4.06$ the planet properties of the two transient objects to the KOI are not heavily affected.

\subsubsection{KOI-0703}

This is the faintest close companion detected in our sample. A less than $3-\sigma$ detection has been achieved for this companion so that results for this object may be flagged by this fact. 

\subsubsection{KOI-1375 \& KOI-0387}

A very close visual stellar companion has been detected to these KOIs. In the case of KOI-1375, the object seems to be bluer than the KOI suggesting that it is probably a background object. Regarding KOI-0387, its companion has the same $i-z$ color. Hence, an assumption of the same age would lead to a similar spectral type. However, the KOI is several magnitudes brighter and redder than the companion, favoring the background nature for this close companion. Note also that observations of KOI-0387 were performed during a non-photometric night. Very large errors are found due to the faintness and closeness of both objects to their correspondent KOIs.



\section{Conclusions}

We have started a high resolution imaging follow-up for the Kepler sample of planet host candidates. The main goal of this survey is to provide additional constrains for the confirmation of the planetary nature  of these candidates and identify those that are possible false positives. A total amount of 98 KOIs (out of the 997) from the second release of the Kepler Team have been properly observed by using the Lucky Imaging technique with the AstraLux instrument at the 2.2m telescope at Calar Alto Observatory. 

Our main results show that  the 58.2\% of the KOIs are actually isolated in terms of not having any visual or bounded companion at less than 6 arcsec. In other words, the 41.9\% of the candidates present close objects. This is an important result in terms of: (a) False positive rate determination, since it points directly which objects have stellar companions that can be mimicking a planet transit detected by Kepler, being then the highest priority for a deeper follow-up with ground-based telescopes to determine the nature of this transit; (b) Updating the planet properties, since as we have shown in the \textit{Discussion} section, they depend on the brightness of the host star. We warn that orbital and physical parameters of the 23 planets orbiting the 17 KOIs with close stellar companions should be revised. (c) Estimating the binary rate in planet host stars. According to their position on a $M_i$ vs. $i-z$ color-magnitude diagram, we have shown that between 6 and 9 of the close companions could be actually bounded to the host star due to their position over our empirical ZAMS. Their distances agree with an S-type binary although still more observations are necessary to confirm both the planet and the binary in all cases. KOIs 0379B, 0658B, 0641B, 0645B and 0298B clearly lie over the ZAMS which suggests a simultaneous formation together with the primary star. Moreover, KOIs 0433B, 0401B, 0592B and 0703B could also lie inside the error bars of the ZAMS but more work should be done to confirm this result. If confirmed, it would imply a lower limit on the observed binary frequency of 6.2-9.2\% Regarding the medium-distance companions (3-6 arcsec), we concluded that only one of them (KOI-0623B) is possibly bounded although we should flag this result due to the large errors in the distance estimates. 

Finally, we have provided accurate astrometric positions and $i$ magnitudes for the close and medium distance companions which could be used to re-compute planet-star parameters in those KOIs affected by the light of the companion. These results add more constrains for theoretical works regarding false positive probabilities for the particular objects studied in this paper.

\begin{acknowledgements}
      This research has been funded by Spanish grants AYA 2010-21161-C02-02, CDS2006-00070 and PRICIT-S2009/ESP-1496. J. Lillo-Box thanks the CSIC JAE-predoc program for the PhD fellowship. H. Bouy is funded by the the Ram\'on y Cajal fellowship program number RYC-2009-04497.
We appreciate the data-sharing and open discussions  with the Kepler Science Team, specially Natalie Batalha and David Ciardi. After we initiated this program in 2011 and contacted them, they have been very cooperative with our efforts, with the general goal of maximize the output of the Kepler mission, in a truly scientific and generous behaviour. We also thank Calar Alto Observatory, and both the open TAC and Spanish GTO panel, for allocation of our observing runs and Felix Hormuth for providing such a useful information about the AstraLux instrument. 

\end{acknowledgements}

\bibliographystyle{aa} 
\bibliography{biblio2.bib} 

\begin{thebibliography}{33}
\expandafter\ifx\csname natexlab\endcsname\relax\def\natexlab#1{#1}\fi

\bibitem[{{Batalha} {et~al.}(2012){Batalha}, {Rowe}, {Bryson}, {Barclay},
  {Burke}, {Caldwell}, {Christiansen}, {Mullally}, {Thompson}, {Brown},
  {Dupree}, {Fabrycky}, {Ford}, {Fortney}, {Gilliland}, {Isaacson}, {Latham},
  {Marcy}, {Quinn}, {Ragozzine}, {Shporer}, {Borucki}, {Ciardi}, {Gautier},
  {Haas}, {Jenkins}, {Koch}, {Lissauer}, {Rapin}, {Basri}, {Boss}, {Buchhave},
  {Charbonneau}, {Christensen-Dalsgaard}, {Clarke}, {Cochran}, {Demory},
  {Devore}, {Esquerdo}, {Everett}, {Fressin}, {Geary}, {Girouard}, {Gould},
  {Hall}, {Holman}, {Howard}, {Howell}, {Ibrahim}, {Kinemuchi}, {Kjeldsen},
  {Klaus}, {Li}, {Lucas}, {Morris}, {Prsa}, {Quintana}, {Sanderfer},
  {Sasselov}, {Seader}, {Smith}, {Steffen}, {Still}, {Stumpe}, {Tarter},
  {Tenenbaum}, {Torres}, {Twicken}, {Uddin}, {Van Cleve}, {Walkowicz}, \&
  {Welsh}}]{batalha12}
{Batalha}, N.~M., {Rowe}, J.~F., {Bryson}, S.~T., {et~al.} 2012, ArXiv e-prints

\bibitem[{{Bayo} {et~al.}(2008){Bayo}, {Rodrigo}, {Barrado Y Navascu{\'e}s},
  {Solano}, {Guti{\'e}rrez}, {Morales-Calder{\'o}n}, \& {Allard}}]{bayo08}
{Bayo}, A., {Rodrigo}, C., {Barrado Y Navascu{\'e}s}, D., {et~al.} 2008, \aap,
  492, 277

\bibitem[{{Bertin} \& {Arnouts}(1996)}]{bertin96}
{Bertin}, E. \& {Arnouts}, S. 1996, \aaps, 117, 393

\bibitem[{{Borucki} {et~al.}(2011){Borucki}, {Koch}, {Basri}, {Batalha},
  {Brown}, {Bryson}, {Caldwell}, {Christensen-Dalsgaard}, {Cochran}, {DeVore},
  {Dunham}, {Gautier}, {Geary}, {Gilliland}, {Gould}, {Howell}, {Jenkins},
  {Latham}, {Lissauer}, {Marcy}, {Rowe}, {Sasselov}, {Boss}, {Charbonneau},
  {Ciardi}, {Doyle}, {Dupree}, {Ford}, {Fortney}, {Holman}, {Seager},
  {Steffen}, {Tarter}, {Welsh}, {Allen}, {Buchhave}, {Christiansen}, {Clarke},
  {Das}, {D{\'e}sert}, {Endl}, {Fabrycky}, {Fressin}, {Haas}, {Horch},
  {Howard}, {Isaacson}, {Kjeldsen}, {Kolodziejczak}, {Kulesa}, {Li}, {Lucas},
  {Machalek}, {McCarthy}, {MacQueen}, {Meibom}, {Miquel}, {Prsa}, {Quinn},
  {Quintana}, {Ragozzine}, {Sherry}, {Shporer}, {Tenenbaum}, {Torres},
  {Twicken}, {Van Cleve}, {Walkowicz}, {Witteborn}, \& {Still}}]{borucki11}
{Borucki}, W.~J., {Koch}, D.~G., {Basri}, G., {et~al.} 2011, \apj, 736, 19

\bibitem[{Brandner {et~al.}(2000)Brandner, Zinnecker, Alcal{\'a}, Allard,
  Covino, Frink, K{\"o}hler, Kunkel, Moneti, \& Schweitzer}]{brandner00}
Brandner, W., Zinnecker, H., Alcal{\'a}, J.~M., {et~al.} 2000, The Astronomical
  Journal, Volume 120, Issue 2, pp. 950-962.

\bibitem[{{Cutri} {et~al.}(2003){Cutri}, {Skrutskie}, {van Dyk}, {Beichman},
  {Carpenter}, {Chester}, {Cambresy}, {Evans}, {Fowler}, {Gizis}, {Howard},
  {Huchra}, {Jarrett}, {Kopan}, {Kirkpatrick}, {Light}, {Marsh}, {McCallon},
  {Schneider}, {Stiening}, {Sykes}, {Weinberg}, {Wheaton}, {Wheelock}, \&
  {Zacarias}}]{cutri03}
{Cutri}, R.~M., {Skrutskie}, M.~F., {van Dyk}, S., {et~al.} 2003, VizieR Online
  Data Catalog, 2246, 0

\bibitem[{{Daemgen} {et~al.}(2009){Daemgen}, {Hormuth}, {Brandner}, {Bergfors},
  {Janson}, {Hippler}, \& {Henning}}]{daemgen09}
{Daemgen}, S., {Hormuth}, F., {Brandner}, W., {et~al.} 2009, \aap, 498, 567

\bibitem[{{Desert} {et~al.}(2011){Desert}, {Charbonneau}, {Fressin}, {Ballard},
  \& {the Kepler Team}}]{desert11}
{Desert}, J.-M., {Charbonneau}, D., {Fressin}, F., {Ballard}, S., \& {the
  Kepler Team}. 2011, American Astronomical Society, ESS meeting \#2, \#4.03,
  2, 403

\bibitem[{{Desidera} \& {Barbieri}(2007)}]{desidera07}
{Desidera}, S. \& {Barbieri}, M. 2007, \aap, 462, 345

\bibitem[{{Doyle} {et~al.}(2011){Doyle}, {Carter}, {Fabrycky}, {Slawson},
  {Howell}, {Winn}, {Orosz}, {Welsh}, {Quinn}, {Latham}, {Torres}, {Buchhave},
  {Marcy}, {Fortney}, {Shporer}, {Ford}, {Lissauer}, {Ragozzine}, {Rucker},
  {Batalha}, {Jenkins}, {Borucki}, {Koch}, {Middour}, {Hall}, {McCauliff},
  {Fanelli}, {Quintana}, {Holman}, {Caldwell}, {Still}, {Stefanik}, {Brown},
  {Esquerdo}, {Tang}, {Furesz}, {Geary}, {Berlind}, {Calkins}, {Short},
  {Steffen}, {Sasselov}, {Dunham}, {Cochran}, {Boss}, {Haas}, {Buzasi}, \&
  {Fischer}}]{doyle11}
{Doyle}, L.~R., {Carter}, J.~A., {Fabrycky}, D.~C., {et~al.} 2011, Science,
  333, 1602

\bibitem[{{Duch{\^e}ne}(1999)}]{duchene99}
{Duch{\^e}ne}, G. 1999, \aap, 341, 547

\bibitem[{{Eggenberger} {et~al.}(2004){Eggenberger}, {Udry}, \&
  {Mayor}}]{eggenberg04}
{Eggenberger}, A., {Udry}, S., \& {Mayor}, M. 2004, \aap, 417, 353

\bibitem[{{Holman} \& {Wiegert}(1999)}]{holman99}
{Holman}, M.~J. \& {Wiegert}, P.~A. 1999, \aj, 117, 621

\bibitem[{{Jenkins} {et~al.}(2010){Jenkins}, {Caldwell}, {Chandrasekaran},
  {Twicken}, {Bryson}, {Quintana}, {Clarke}, {Li}, {Allen}, {Tenenbaum}, {Wu},
  {Klaus}, {Middour}, {Cote}, {McCauliff}, {Girouard}, {Gunter}, {Wohler},
  {Sommers}, {Hall}, {Uddin}, {Wu}, {Bhavsar}, {Van Cleve}, {Pletcher},
  {Dotson}, {Haas}, {Gilliland}, {Koch}, \& {Borucki}}]{jenkins10}
{Jenkins}, J.~M., {Caldwell}, D.~A., {Chandrasekaran}, H., {et~al.} 2010,
  \apjl, 713, L87

\bibitem[{{Kley}(2010)}]{kley10}
{Kley}, W. 2010, in EAS Publications Series, Vol.~42, EAS Publications Series,
  ed. {K.~Go{\.z}dziewski, A.~Niedzielski, \& J.~Schneider}, 227--238

\bibitem[{{Law} {et~al.}(2006){Law}, {Mackay}, \& {Baldwin}}]{law06}
{Law}, N.~M., {Mackay}, C.~D., \& {Baldwin}, J.~E. 2006, \aap, 446, 739

\bibitem[{{Marcy} \& {Butler}(1996)}]{marcy96}
{Marcy}, G.~W. \& {Butler}, R.~P. 1996, \apjl, 464, L147

\bibitem[{{Mayor} \& {Queloz}(1995)}]{mayor95}
{Mayor}, M. \& {Queloz}, D. 1995, \nat, 378, 355

\bibitem[{{Monet} {et~al.}(2003){Monet}, {Levine}, {Canzian}, {Ables}, {Bird},
  {Dahn}, {Guetter}, {Harris}, {Henden}, {Leggett}, {Levison}, {Luginbuhl},
  {Martini}, {Monet}, {Munn}, {Pier}, {Rhodes}, {Riepe}, {Sell}, {Stone},
  {Vrba}, {Walker}, {Westerhout}, {Brucato}, {Reid}, {Schoening}, {Hartley},
  {Read}, \& {Tritton}}]{monet03}
{Monet}, D.~G., {Levine}, S.~E., {Canzian}, B., {et~al.} 2003, \aj, 125, 984

\bibitem[{{Moraux} {et~al.}(2003){Moraux}, {Bouvier}, {Stauffer}, \&
  {Cuillandre}}]{moraux03}
{Moraux}, E., {Bouvier}, J., {Stauffer}, J.~R., \& {Cuillandre}, J.-C. 2003,
  \aap, 400, 891

\bibitem[{{Morton} \& {Johnson}(2011)}]{timothy11}
{Morton}, T.~D. \& {Johnson}, J.~A. 2011, \apj, 738, 170

\bibitem[{{Mugrauer} \& {Neuh{\"a}user}(2009)}]{mugrauer09}
{Mugrauer}, M. \& {Neuh{\"a}user}, R. 2009, \aap, 494, 373

\bibitem[{{O'Donovan} {et~al.}(2006){O'Donovan}, {Charbonneau}, {Torres},
  {Mandushev}, {Dunham}, {Latham}, {Alonso}, {Brown}, {Esquerdo}, {Everett}, \&
  {Creevey}}]{odonovan06}
{O'Donovan}, F.~T., {Charbonneau}, D., {Torres}, G., {et~al.} 2006, \apj, 644,
  1237

\bibitem[{{Ofek}(2008)}]{ofek08}
{Ofek}, E.~O. 2008, \pasp, 120, 1128

\bibitem[{{Pickles}(1998)}]{pickles98}
{Pickles}, A.~J. 1998, VizieR Online Data Catalog, 611, 863

\bibitem[{{Raghavan} {et~al.}(2010){Raghavan}, {McAlister}, {Henry}, {Latham},
  {Marcy}, {Mason}, {Gies}, {White}, \& {ten Brummelaar}}]{raghavan10}
{Raghavan}, D., {McAlister}, H.~A., {Henry}, T.~J., {et~al.} 2010, \apjs, 190,
  1

\bibitem[{Schmidt-Kaler(1982)}]{schmidt-kaler82}
Schmidt-Kaler, T. 1982, Springer-Verlag, Vol. Vol. 2B, Physical parameters of
  the stars (New York: Landolt-B\"ornstein New Series)

\bibitem[{{Seager} \& {Mall{\'e}n-Ornelas}(2003)}]{seager03}
{Seager}, S. \& {Mall{\'e}n-Ornelas}, G. 2003, \apj, 585, 1038

\bibitem[{{Smith} {et~al.}(2002){Smith}, {Tucker}, {Kent}, {Richmond},
  {Fukugita}, {Ichikawa}, {Ichikawa}, {Jorgensen}, {Uomoto}, {Gunn}, {Hamabe},
  {Watanabe}, {Tolea}, {Henden}, {Annis}, {Pier}, {McKay}, {Brinkmann}, {Chen},
  {Holtzman}, {Shimasaku}, \& {York}}]{smith02}
{Smith}, J.~A., {Tucker}, D.~L., {Kent}, S., {et~al.} 2002, \aj, 123, 2121

\bibitem[{{Stello} \& {Nissen}(2001)}]{stello01}
{Stello}, D. \& {Nissen}, P.~E. 2001, \aap, 374, 105

\bibitem[{{Strehl}(1902)}]{strehl1902}
{Strehl}, K. 1902, Astronomische Nachrichten, 158, 89

\bibitem[{{Thebault}(2011)}]{thebault11}
{Thebault}, P. 2011, Celestial Mechanics and Dynamical Astronomy, 111, 29

\bibitem[{{Yanny} {et~al.}(1994){Yanny}, {Guhathakurta}, {Bahcall}, \&
  {Schneider}}]{yanny94}
{Yanny}, B., {Guhathakurta}, P., {Bahcall}, J.~N., \& {Schneider}, D.~P. 1994,
  \aj, 107, 1745

\end{thebibliography}

\clearpage

\setcounter{table}{0} 

\begin{table}
\caption{Plate solution for our photometric observations}             
\label{astrometry}      
\centering                          
\begin{tabular}{c c c}        
\hline\hline                 
Coeff. & $\xi$ & $\eta$ \\    
\hline                        
$c_{10}$ & $-23.58 $ $mas/px$ & -0.65 $mas/px$ \\
$c_{01}$ & $-0.81 $ $mas/px$ & 23.58 $mas/px$\\
$c_{20}$ & $4.05\times10^{-3}$ $mas/px^2$ &  $7.93 \times 10^{-5}$ $mas/px^2$ \\
$c_{02}$ & $1.31\times10^{-3}$ $mas/px^2$ & $9.35 \times 10^{-5}$ $mas/px^2$\\ \hline
rms & 48.5 mas & 34.5 mas \\ \hline
PA & $1.78^{\circ}$ & \\
Pixel scale & 23.59 mas/px & \\


\hline                                   
\end{tabular}
\end{table}
%


\setcounter{table}{1} 

\begin{table}
\caption{Cumulative percentage of KOIs with visual (even bounded or not) and bounded companions according to our observations}             
\setlength{\extrarowheight}{5pt}
\label{mainresults}      
\centering                          
\begin{tabular}{|r | c c c|}        
\hline
 & Within 3''  & Within 6'' & Within 10'' \\    
\hline                        
Visual & 17.3 \%  & 41.8 \%  & 66.3 \% \\
Bounded &  6.1-9.2 \% & 6.1-9.2 \% & --- \\ 

\hline                                   
\end{tabular}
\tablefoot{\textit{Visual} means here any kind of stellar source detected around the KOI, even bounded or not. The \textit{Bounded} sample refers to those objects probably bounded according to the study performed in section \S~\ref{boundedornot}.}
\end{table}
%


\begin{table}
\caption{Estimated spectral types for objects with $i-z>0.21$}             
\label{companionspt}      
\centering                          
\begin{tabular}{c c c c}        
\hline\hline                 
Object & Lower Limit SpT & SpT & Upper Limit SpT \\    
\hline                        
   KOI-0641A     &     K5      &    K7      &    K9     \\
   KOI-0641B      &    M1     &     M3     &     M3 \\
   KOI-0641C     &     M3     &     M5    &      M5 \\
   KOI-0703B      &    K9      &    M1      &    M3 \\
   KOI-0298A      &    K5      &    K7      &    K9 \\
   KOI-0298B      &    K9     &     M0      &    M2 \\
   KOI-0658B      &    M3     &     M3     &     M3 \\
   KOI-0401B      &    K9      &    K9      &    M1 \\
   KOI-0422B      &    M5      &    M5     &     M6 \\
   KOI-0645B      &    K9      &    K9      &    M1 \\
   KOI-0433B      &    M3     &     M5      &    M6 \\
\hline                                   
\end{tabular}
\tablefoot{Second and third columns represent the lower and upper limits (respectively) for the spectral type determination by the method described in the text. Since no decimals have been considered, we estimate 1 subclass error for those with no differences between the error limits and the central value.}
\end{table}
%


\begin{table*}
\caption{Observing information of the 41 non-isolated objects in our sample.}             
\label{obsresumen}      
\centering                          
\begin{tabular}{cccccrccccc}        
\hline\hline                 
KOI & RA\tablefootmark{a} & DEC\tablefootmark{a} & Date  & \centering{ExpTime}\tablefootmark{b} & Filter &  $i_{Complete}$ & $i_{Detect}$\tablefootmark{c}\\
ID & J2000.0 & J2000.0 & &   \centering{$s$} & & $mag$ & $mag$\\
\hline                        

  99 & 19:41:44.23 & +44:31:52.0 & 05jul11 & 60.0 & i & 16.34 & 19.94\\
  131 & 19:56:23.42 & +43:29:51.4 & 10jul11 & 150.0 & i & 17.34 & 20.94\\
  212 & 19:44:33.54 & +41:36:11.5 & 09jul11 & 240.0 & i & 17.65 & 21.25\\
  232 & 19:24:26.86 & +39:56:56.8 & 02jul11 & 200.0 & i & 18.40 & 22.00\\
  238 & 19:47:59.68 & +42:46:55.2 & 05jul11 & 200.0 & i & 17.65 & 21.25\\
  298 & 19:21:58.61 & +52:03:19.8 & 12jun11 & 160.0 & i & 17.41 & 21.01\\
  298 & 19:21:58.61 & +52:03:19.8 & 12jun11 & 200.0 & z         & --- & --- \\
  326 & 19:06:37.44 & +46:47:00.6 & 12jun11 & 200.0 & i & 18.40 & 22.00\\
  343 & 19:40:28.52 & +48:28:52.7 & 12jun11 & 200.0 & i & 18.40 & 22.00\\
  372 & 19:56:29.40 & +41:52:00.5 & 01jul11 & 34.0 & i & 16.48 & 20.08\\
  372 & 19:56:29.40 & +41:52:00.5 & 01jul11 & 40.0 & z         & --- & ---\\
  375 & 19:24:48.28 & +51:08:39.5 & 02jul11 & 87.0 & i & 17.50 & 21.10\\
  379 & 19:28:13.62 & +37:46:34.3 & 25jul11 & 140.0 & i & 17.65 & 21.25\\
  379 & 19:28:13.62 & +37:46:34.3 & 25jul11 & 140.0 & z         & --- & ---\\
  387 & 19:08:52.48 & +38:51:45.0 & 02jul11 & 90.0 & i & 17.53 & 21.13\\
  387 & 19:08:52.48 & +38:51:45.0 & 02jul11 & 100.0 & z         & --- & ---\\
  401 & 19:03:24.88 & +38:23:02.8 & 05jul11 & 200.0 & i & 18.40 & 22.00\\
  401 & 19:03:24.88 & +38:23:02.8 & 05jul11 & 200.0 & z         & --- & ---\\
  433 & 19:54:12.20 & +48:19:57.0 & 07jul11 & 200.0 & i & 17.65 & 21.25\\
  433 & 19:54:12.20 & +48:19:57.0 & 07jul11 & 200.0 & z         & --- & ---\\
  439 & 19:45:37.66 & +51:21:29.5 & 11jun11 & 200.0 & i & 18.40 & 22.00\\
  465 & 19:35:42.83 & +45:08:33.0 & 10jul11 & 200.0 & i & 17.65 & 21.25\\
  520 & 19:38:40.31 & +43:51:11.9 & 08jul11 & 200.0 & i & 18.40 & 22.00\\
  548 & 19:18:00.18 & +51:41:08.5 & 09jun11 & 200.0 & i & 18.40 & 22.00\\
  555 & 19:32:29.62 & +40:56:05.3 & 10jun11 & 200.0 & i & 18.40 & 22.00\\
  592 & 19:37:51.02 & +46:49:17.4 & 07jul11 & 200.0 & i & 17.65 & 21.25\\
  592 & 19:37:51.02 & +46:49:17.4 & 08jul11 & 200.0 & z         & --- & ---\\
  611 & 19:53:10.57 & +41:41:01.7 & 02jul11 & 200.0 & i & 18.40 & 22.00\\
  623 & 19:40:54.34 & +50:33:32.4 & 05jul11 & 30.0 & i & 16.34 & 19.94\\
  626 & 19:40:46.42 & +39:32:22.9 & 04jul11 & 180.0 & i & 18.09 & 21.69\\
  626 & 19:40:46.42 & +39:32:22.9 & 04jul11 & 180.0 & z         & --- & ---\\
  628 & 19:14:47.69 & +39:42:29.9 & 04jul11 & 200.0 & i & 18.40 & 22.00\\
  628 & 19:14:47.69 & +39:42:29.9 & 04jul11 & 160.0 & z         & --- & ---\\
  638 & 19:42:14.26 & +40:14:10.7 & 05jul11 & 140.0 & i & 17.65 & 21.25\\
  641 & 19:57:11.88 & +40:14:06.4 & 01jul11 & 87.0 & i & 17.50 & 21.10\\
  641 & 19:57:11.88 & +40:14:06.4 & 01jul11 & 87.0 & z         & --- & ---\\
  644 & 19:19:52.03 & +40:31:57.7 & 04jul11 & 180.0 & i & 18.09 & 21.69\\
  644 & 19:19:52.03 & +40:31:57.7 & 04jul11 & 180.0 & z         & --- & ---\\
  645 & 19:40:52.18 & +40:35:32.3 & 26jun11 & 200.0 & i & 17.65 & 21.25\\
  645 & 19:40:52.18 & +40:35:32.3 & 26jun11 & 200.0 & z         & --- & ---\\
  658 & 19:48:21.60 & +41:23:16.8 & 01jul11 & 200.0 & i & 18.40 & 22.00\\
  658 & 19:48:21.60 & +41:23:16.8 & 01jul11 & 200.0 & z         & --- & ---\\
  685 & 19:41:54.17 & +43:29:35.2 & 11jun11 & 200.0 & z         & --- & ---\\
  685 & 19:41:54.17 & +43:29:35.2 & 11jun11 & 200.0 & i & 18.40 & 22.00\\
  703 & 19:39:38.88 & +45:34:00.1 & 11jun11 & 60.0 & i & 16.89 & 20.49\\
  703 & 19:39:38.88 & +45:34:00.1 & 11jun11 & 200.0 & z         & --- & ---\\
  704 & 18:57:32.69 & +45:43:10.9 & 05jul11 & 150.0 & i & 17.65 & 21.25\\
  721 & 19:48:16.42 & +46:50:03.5 & 26jun11 & 140.0 & i & 17.65 & 21.25\\
  721 & 19:48:16.42 & +46:50:03.5 & 26jun11 & 140.0 & z         & --- & ---\\
  841 & 19:28:56.82 & +41:05:09.2 & 11jun11 & 200.0 & i & 18.40 & 22.00\\
  841 & 19:28:56.82 & +41:05:09.2 & 11jun11 & 200.0 & z         & --- & ---\\
  881 & 19:39:38.34 & +42:56:07.1 & 11jun11 & 200.0 & i & 18.40 & 22.00\\
  1032 & 19:27:54.61 & +37:31:57.4 & 06jul11 & 150.0 & i & 18.09 & 21.69\\
  1192 & 19:24:07.70 & +38:42:14.0 & 06jul11 & 200.0 & i & 17.65 & 21.25\\
  1375 & 19:13:16.90 & +42:15:41.0 & 06jul11 & 150.0 & i & 17.34 & 20.94\\
  1375 & 19:13:16.90 & +42:15:41.0 & 06jul11 & 150.0 & z         & --- & ---\\
  1527 & 19:46:41.12 & +43:29:54.2 & 07jul11 & 200.0 & i & 17.41 & 21.01\\
  1573 & 19:47:23.06 & +40:08:19.0 & 08jul11 & 200.0 & i & 18.40 & 22.00\\
  1574 & 19:51:40.07 & +46:57:54.4 & 06jul11 & 200.0 & i & 17.65 & 21.25\\
\hline                                   
\end{tabular}
\tablefoot{ \\
\tablefoottext{a}{Right ascension and Declination from \cite{borucki11}}\\
\tablefoottext{b}{Effective exposure time of the image. As the selection rate was of 10\% for all the images, one must multiply this column by 10 to obtain the real exposure time.}\\
\tablefoottext{c}{Estimated completeness and detectability magnitudes scaled to the ones found for the globular cluster M15 (see section \S~\ref{sensitivity}) by the exposure time of each particular image }
}
\end{table*}


\longtabL{5}{
\begin{landscape}
\begin{longtable}{rlcrccccccc}
\caption{\label{results} Photometric and astrometric results for companions closer than 3 arcsec detected in our KOI sample (19 companions to 17 KOIs)}\\
\hline\hline
KOI & Comp. & Angular Sep.  & Angle (deg) & Phot? & SDSSi & SDSSz & $\Delta i$ & $\Delta z$ &$(i-z)$ & $(i-z)$\_unc \\
        &  & arcsec & degrees &   & mag &  mag & mag & mag &mag & mag \\
\hline
\endfirsthead
\caption{continued.}\\
\hline\hline
KOI & Companion & Angular Sep. & Angle (deg) & Phot? & SDSSi & SDSSz & $\Delta i$ & $\Delta z$ &$(i-z)$ & $(i-z)$\_unc \\
        &  & arcsec & degrees &   & mag &  mag & mag & mag &mag & mag \\
  \hline
\endhead
\hline
\endfoot
  298  & A* &   0.000 &    0.0 & P & 13.16 $\pm$ 0.15 & 12.88 $\pm$  0.16  & 0.0 $\pm$ 0.12  & 0.0 $\pm$  0.16  & 0.273 $\pm$ 0.073   & 0.22  \\   
       & B+ & 1.963   & 273.09 & P & 13.64 $\pm$ 0.18 & 13.23 $\pm$   0.20 & 0.48 $\pm$ 0.16 & 0.35 $\pm$   0.2 & 0.405 $\pm$ 0.073   & 0.34  \\   
       & C  & 1.961   &  93.56 & P & 17.59 $\pm$ 0.53 & ---                & 4.43 $\pm$ 0.53 & ---              & ---                 & ---  \\  
  379  & A* &   0.000 &    0.0 & N & 13.29 $\pm$ 0.07 & 13.22 $\pm$  0.08  & 0.0 $\pm$ 0.05  & 0.0 $\pm$  0.06  & 0.07 $\pm$ 0.047    & 0.11  \\   
       & B+ & 1.898   &  80.44 & N & 14.25 $\pm$ 0.09 & 14.10 $\pm$  0.09  & 0.96 $\pm$ 0.07 & 0.88 $\pm$  0.08 & 0.15 $\pm$ 0.048    & 0.15  \\   
  387  & A  &   0.000 &    0.0 & N & 12.80 $\pm$ 0.25 & 12.79 $\pm$  0.22  & 0.0 $\pm$ 0.35  & 0.0 $\pm$  0.28  & 0.0080 $\pm$ 0.101  & 0.33  \\   
       & B  & 0.894   & 350.78 & N & 16.05 $\pm$ 0.96 & 16.05 $\pm$  0.75  & 3.25 $\pm$ 0.99 & 3.26 $\pm$  0.77 & -0.0040 $\pm$ 0.102 & 1.3  \\    
  401  & A  &   0.000 &    0.0 & Y & 13.76 $\pm$ 0.07 & 13.69 $\pm$  0.07  & 0.0 $\pm$ 0.04  & 0.0 $\pm$  0.06  & 0.07 $\pm$ 0.049    & 0.1  \\    
       & B+ & 1.946   & 269.98 & Y & 16.35 $\pm$ 0.21 & 15.97 $\pm$  0.22  & 2.59 $\pm$  0.2 & 2.27 $\pm$  0.22 & 0.383 $\pm$ 0.049   & 0.32  \\   
  433  & A  &   0.000 &    0.0 & Y & 14.67 $\pm$ 0.19 & 14.54 $\pm$  0.19  & 0.0 $\pm$ 0.18  & 0.0 $\pm$  0.18  & 0.13 $\pm$  0.27    & 0.27  \\   
       & B+ & 2.326   &   5.64 & Y & 18.13 $\pm$ 0.19 & 17.52 $\pm$  0.19  & 4.0 $\pm$ 0.18  & 2.98 $\pm$  0.18 & 0.89 $\pm$  0.29    & 0.27  \\   
  592  & A  &   0.000 &    0.0 & Y & 14.13 $\pm$ 0.07 & 14.57 $\pm$  0.06  & 0.0 $\pm$  0.0  & 0.0 $\pm$ 0.0    & -0.444 $\pm$ 0.076  & 0.09  \\   
       & B  & 2.264   &  151.3 & Y & 17.94 $\pm$ 0.36 & 18.02 $\pm$  0.39  & 3.81 $\pm$ 0.35 & 3.46 $\pm$ 0.38  & -0.094 $\pm$ 0.522  & ---  \\  
  626  & A  &   0.000 &    0.0 & N & 13.36 $\pm$ 0.07 & 13.37 $\pm$  0.09  & 0.0 $\pm$  0.0  & 0.0 $\pm$   0.0  & -0.01 $\pm$ 0.081   & 0.11  \\   
       & B  & 2.694   & 349.02 & N & 18.30 $\pm$ 0.36 & 18.34 $\pm$  0.41  & 4.95 $\pm$ 0.36 & 4.97 $\pm$   0.4 & -0.036 $\pm$ 0.086  & 0.55  \\   
  628  & A  &   0.000 &    0.0 & N & 13.69 $\pm$ 0.07 & 13.75 $\pm$  0.09  & 0.0 $\pm$  0.0  & 0.0 $\pm$   0.0  & -0.061 $\pm$ 0.081  & 0.11  \\   
       & B  & 1.765   & 311.42 & N & 17.43 $\pm$ 0.33 & 17.41 $\pm$  0.37  & 3.75 $\pm$ 0.32 & 3.67 $\pm$  0.36 & 0.018 $\pm$ 0.082   & 0.5  \\    
       & C  & 2.695   & 239.68 & N & 17.84 $\pm$ 0.33 & 17.79 $\pm$  0.37  & 4.16 $\pm$ 0.32 & 4.05 $\pm$  0.36 & 0.13 $\pm$ 0.084    & 0.69  \\   
  641  & A* &   0.000 &    0.0 & P & 13.25 $\pm$ 0.12 & 12.99 $\pm$  0.13  & 0.0 $\pm$ 0.02  & 0.0 $\pm$  0.08  & 0.265 $\pm$ 0.087   & 0.17  \\   
       & B+ & 2.206   & 277.22 & P & 14.90 $\pm$ 0.12 & 14.36 $\pm$  0.19  & 1.65 $\pm$ 0.03 & 1.37 $\pm$  0.16 & 0.54 $\pm$ 0.087    & 0.23  \\   
  644  & A  &   0.000 &    0.0 & N & 13.43 $\pm$ 0.07 & 13.31 $\pm$  0.09  & 0.0 $\pm$  0.0  & 0.0 $\pm$   0.0  & 0.123 $\pm$ 0.081   & 0.11  \\   
       & B  & 2.636   &  63.63 & N & 14.87 $\pm$ 0.08 & 14.73 $\pm$  0.09  & 1.44 $\pm$ 0.04 & 1.43 $\pm$  0.04 & 0.136 $\pm$ 0.081   & 0.12  \\   
  645  & A  &   0.000 &    0.0 & Y & 13.55 $\pm$ 0.06 & 13.51 $\pm$  0.06  & 0.0 $\pm$  0.0  & 0.0 $\pm$   0.0  & 0.043 $\pm$ 0.038   & 0.08  \\   
       & B+ &  2.78   &  50.83 & Y & 15.72 $\pm$ 0.08 & 15.33 $\pm$  0.09  & 2.17 $\pm$ 0.05 & 1.82 $\pm$  0.06 & 0.387 $\pm$ 0.038   & 0.11  \\   
  658  & A  &   0.000 &    0.0 & P & 13.74 $\pm$ 0.06 & 13.72 $\pm$  0.06  & 0.0 $\pm$  0.0  & 0.0 $\pm$   0.0  & 0.018 $\pm$ 0.064   & 0.08  \\   
       & B+ & 1.865   & 240.08 & P & 18.38 $\pm$ 0.29 & 17.79 $\pm$   0.30 & 4.63 $\pm$ 0.29 & 4.06 $\pm$   0.3 & 0.588 $\pm$ 0.067   & 0.42  \\   
  703  & A  &   0.000 &    0.0 & P & 13.01 $\pm$ 0.11 & 13.06 $\pm$  0.12  & 0.0 $\pm$  0.0  & 0.0 $\pm$   0.0  & -0.052 $\pm$ 0.087  & 0.16  \\   
       & B+ &  1.89   &  34.44 & P & 19.50 $\pm$ 0.41 & 19.04 $\pm$   0.6  & 6.48 $\pm$  0.4 & 5.98 $\pm$  0.59 & 0.452 $\pm$ 0.112   & 0.73  \\   
  704  & A  &   0.000 &    0.0 & Y & 13.45 $\pm$ 0.06 & ---                & 0.0 $\pm$  0.0  & ---              & ---                 & ---  \\  
       & B  & 1.657   & 180.36 & Y & 18.47 $\pm$ 0.61 & ---                & 5.02 $\pm$  0.6 & ---              & ---                 & ---  \\  
  721  & A  &   0.000 &    0.0 & Y & 13.52 $\pm$ 0.06 & 13.44 $\pm$  0.06  & 0.0 $\pm$ 0.01  & 0.0 $\pm$  0.03  & 0.075 $\pm$ 0.038   & 0.08  \\   
       & B  & 1.844   & 195.67 & Y & 17.37 $\pm$ 0.13 & 17.28 $\pm$  0.14  & 3.85 $\pm$ 0.12 & 3.84 $\pm$  0.13 & 0.092 $\pm$  0.04   & 0.2  \\    
  841  & A  &   0.000 &    0.0 & P & 15.57 $\pm$ 0.26 & 15.52 $\pm$  0.26  & 0.0 $\pm$ 0.18  & 0.0 $\pm$  0.18  & 0.05 $\pm$  0.37    & 0.37  \\   
       & B  & 1.956   &  69.13 & P & 19.25 $\pm$ 0.26 & 19.16 $\pm$  0.26  & 3.68 $\pm$ 0.18 & 3.64 $\pm$  0.19 & -0.01 $\pm$  0.42   & 0.37  \\   
  1375 & A  &   0.000 &    0.0 & Y & 13.24 $\pm$ 0.23 & 13.29 $\pm$  0.19  & 0.0 $\pm$ 0.32  & 0.0 $\pm$  0.25  & -0.051 $\pm$ 0.057  & 0.3  \\    
       & B  & 0.796   & 266.46 & Y & 15.99 $\pm$ 0.55 & 16.23 $\pm$  0.46  & 2.75 $\pm$ 0.59 & 2.94 $\pm$  0.49 & -0.241 $\pm$ 0.059  & 0.82  \\     
\end{longtable}
\tablefoot{Objects marked with an asterisk (*) in the \textit{Companion} column are assumed to be the Kepler Object of Interest. Close companions with similar magnitudes are found in these cases being difficult to distinguish which of them is the actual host. The assumed KOI is the brightest object in the system. Companions marked with a '+' symbol indicate those probably bounded according to section ~\ref{closebounded}. The last column shows the photometric error if we take into account the empirical errors of the software gain of the intrument.}
\end{landscape}
}


\longtabL{6}{
\begin{landscape}
\begin{longtable}{rlcrcccccc}
\caption{\label{results36} Photometric and astrometric results for companions between 3-6 arcsec detected in our KOI sample (30 companions to 27 KOIs)}\\
\hline\hline
KOI & Comp. & Angular Sep.  & Angle (deg) & Phot? & SDSSi & SDSSz & $\Delta i$ & $\Delta z$ &$(i-z)$ \\
        &  & arcsec & degrees &   & mag &  mag & mag & mag &mag  \\
\hline
\endfirsthead
\caption{continued.}\\
\hline\hline
KOI & Comp. & Angula Sep. & Angle (deg) & Phot? & SDSSi & SDSSz & $\Delta i$ & $\Delta z$ &$(i-z)$ \\
        &  & arcsec & degrees &   & mag &  mag & mag & mag &mag  \\
  \hline
\endhead
\hline
\endfoot
 99	& A     & 0.000 & 0.0   & Y     & 12.57 $\pm$ 0.18	& ---			& 0.00  $\pm$ 0.18	& ---		  & ---    \\
 99	& B     & 3.383 & 49.3  & Y     & 17.96 $\pm$ 0.18	& ---			& 5.39  $\pm$ 0.19	& ---		  & ---    \\
 99	& C     & 5.997 & 338.3 & Y     & 16.58 $\pm$ 0.18	& ---			& 4.01  $\pm$ 0.18	& ---		  & ---    \\
 131	& A     & 0.000 & 0.0   & Y     & 13.53 $\pm$ 0.17	& ---			& 0.00  $\pm$ 0.18	& ---		  & ---    \\
 131	& B     & 5.540 & 154.6 & Y     & 17.00 $\pm$ 0.17	& ---			& 3.48  $\pm$ 0.18	& ---		  & ---    \\
 212	& A     & 0.000 & 0.0   & P     & 14.70 $\pm$ 0.19	& ---			& 0.00  $\pm$ 0.18	& ---		  & ---    \\
 212	& B     & 4.812 & 308.8 & P     & 17.52 $\pm$ 0.19	& ---			& 2.82  $\pm$ 0.18	& ---		  & ---    \\
 232	& A     & 0.000 & 0.0   & N     & 13.76 $\pm$ 0.16	& ---			& 0.00  $\pm$ 0.18	& ---		  & ---    \\
 232	& B     & 5.550 & 22.3  & N     & 18.19 $\pm$ 0.16	& ---			& 4.43  $\pm$ 0.18	& ---		  & ---    \\
 238	& A     & 0.000 & 0.0   & Y     & 13.74 $\pm$ 0.18	& ---			& 0.00  $\pm$ 0.18	& ---		  & ---    \\
 238	& B     & 3.769 & 328.7 & Y     & 20.61 $\pm$ 0.24	& ---			& 6.87  $\pm$ 0.24	& ---		  & ---    \\
 326	& A     & 0.000 & 0.0   & P     & 14.84 $\pm$ 0.26	& ---			& 0.00  $\pm$ 0.18	& ---		  & ---    \\
 326	& B     & 3.452 & 269.4 & P     & 16.61 $\pm$ 0.26	& ---			& 1.77  $\pm$ 0.18	& ---		  & ---    \\
 343	& A     & 0.000 & 0.0   & P     & 13.06 $\pm$ 0.26	& ---			& 0.00  $\pm$ 0.18	& ---		  & ---    \\
 343	& B     & 4.938 & 145.8 & P     & ---   	        & ---			&  ---			& ---		  & ---    \\
 372	& A     & 0.000 & 0.0   & P     & 12.25 $\pm$ 0.18	& 12.12 $\pm$ 0.18  	& 0.00  $\pm$ 0.18	& 0.00  $\pm$ 0.18& 0.13 $\pm$ 0.25\\ 
 372	& B     & 5.962 & 36.1  & P     & 16.88 $\pm$ 0.18	& 16.58 $\pm$ 0.18  	& 4.64  $\pm$ 0.18	& 4.45  $\pm$ 0.18& 0.30 $\pm$ 0.25\\ 
 375	& A     & 0.000 & 0.0   & N     & 13.08 $\pm$ 0.15	& ---			& 0.00  $\pm$ 0.18	& ---		  & ---    \\
 375	& B     & 5.329 & 156.1 & N     & 18.06 $\pm$ 0.16	& ---			& 4.98  $\pm$ 0.18	& ---		  & ---    \\
 433	& A     & 0.000 & 0.0   & Y     & 14.67 $\pm$ 0.19	& 14.54 $\pm$ 0.19  	& 0.00  $\pm$ 0.18	& 0.00  $\pm$ 0.18& 0.13 $\pm$ 0.27\\ 
 433	& C     & 3.670 & 293.4 & Y     & 17.26 $\pm$ 0.19	& 17.07 $\pm$ 0.19  	& 2.60  $\pm$ 0.18	& 2.53  $\pm$ 0.18& 0.19 $\pm$ 0.27\\ 
 439	& A     & 0.000 & 0.0   & P     & 14.11 $\pm$ 0.32	& ---			& 0.00  $\pm$ 0.18	& ---		  & ---  \\
 439	& B     & 5.453 & 16.9  & P     & 19.45 $\pm$ 0.32	& ---			& 5.34  $\pm$ 0.19	& ---		  & ---  \\
 465	& A     & 0.000 & 0.0   & Y     & 13.94 $\pm$ 0.17	& ---			& 0.00  $\pm$ 0.18	& ---		  & ---  \\
 465	& B     & 3.580 & 130.8 & Y     & 18.30 $\pm$ 0.17	& ---			& 4.35  $\pm$ 0.18	& ---		  & ---  \\
 465	& C     & 4.527 & 193.1 & Y     & 17.82 $\pm$ 0.17	& ---			& 3.87  $\pm$ 0.18	& ---		  & ---  \\
 520	& A     & 0.000 & 0.0   & N     & 14.30 $\pm$ 0.17	& ---			& 0.00  $\pm$ 0.18	& ---		  & ---  \\
 520	& B     & 5.628 & 271.7 & N     & 18.14 $\pm$ 0.17	& ---			& 3.84  $\pm$ 0.18	& ---		  & ---  \\
 548	& A     & 0.000 & 0.0   & N     & 13.37 $\pm$ 0.50	& ---			& 0.00  $\pm$ 0.18	& ---		  & ---  \\
 548	& B     & 4.597 & 133.9 & N     & 18.33 $\pm$ 0.50	& ---			& 4.96  $\pm$ 0.18	& ---		  & ---  \\
 555	& A     & 0.000 & 0.0   & N     & 14.29 $\pm$ 0.18	& ---			& 0.00  $\pm$ 0.18	& ---		  & ---  \\
 555	& B     & 4.021 & 23.1  & N     & 17.86 $\pm$ 0.18	& ---			& 3.58  $\pm$ 0.18	& ---		  & ---  \\
 611	& A     & 0.000 & 0.0   & N     & 13.75 $\pm$ 0.15	& ---			& 0.00  $\pm$ 0.18	& ---		  & ---  \\
 611	& B     & 5.999 & 324.2 & N     & 19.18 $\pm$ 0.16	& ---			& 5.44  $\pm$ 0.18	& ---		  & ---  \\
 623	& A     & 0.000 & 0.0   & Y     & 11.55 $\pm$ 0.18	& ---			& 0.00  $\pm$ 0.18	& ---		  & ---  \\
 623	& B     & 5.540 & 202.0 & Y     & 14.07 $\pm$ 0.18	& ---			& 2.52  $\pm$ 0.18	& ---		  & ---  \\
 638	& A     & 0.000 & 0.0   & Y     & 13.51 $\pm$ 0.17	& ---			& 0.00  $\pm$ 0.18	& ---		  & ---  \\
 638	& B     & 5.938 & 65.5  & Y     & 19.43 $\pm$ 0.19	& ---			& 5.92  $\pm$ 0.19	& ---		  & ---  \\
 641	& A     & 0.000 & 0.0   & P     & 13.25 $\pm$ 0.12	& 12.99 $\pm$ 0.13  	& 0.00  $\pm$ 0.02	& 0.00  $\pm$ 0.08& 0.27 $\pm$ 0.17\\ 
 641	& C     & 3.653 & 206.1 & P     & 13.38 $\pm$ 0.16	& 12.89 $\pm$ 0.17  	& 0.29  $\pm$ 0.18	& 0.08  $\pm$ 0.18& 0.49 $\pm$ 0.23\\ 
 685	& A     & 0.000 & 0.0   & P     & 13.71 $\pm$ 0.3	& ---			& 0.00  $\pm$ 0.18	& ---		  & --- \\
 685	& B     & 3.268 & 271.1 & P     & 19.24 $\pm$ 0.3	& ---			& 5.54  $\pm$ 0.18	& ---		  & --- \\
 841	& A     & 0.000 & 0.0   & P     & 15.57 $\pm$ 0.26	& 15.52 $\pm$ 0.26  	& 0.00  $\pm$ 0.18	& 0.00  $\pm$ 0.18& 0.05 $\pm$ 0.37\\ 
 841	& C     & 5.599 & 41.5  & P     & 17.19 $\pm$ 0.26	& 17.20 $\pm$ 0.26  	& 1.62  $\pm$ 0.18	& 1.68  $\pm$ 0.18& -0.01 $\pm$ 0.37\\ 
 881	& A     & 0.000 & 0.0   & P     & 15.58 $\pm$ 0.26	& ---			& 0.00  $\pm$ 0.18	& ---		  & ---  \\
 881	& B     & 5.264 & 191.5 & P     & 20.70 $\pm$ 0.29	& ---			& 5.12  $\pm$ 0.22	& ---		  & ---  \\
 1032	& A     & 0.000 & 0.0   & Y     & 13.47 $\pm$ 0.17	& ---			& 0.00  $\pm$ 0.18	& ---		  & ---  \\
 1032	& B     & 5.856 & 320.3 & Y     & 18.90 $\pm$ 0.17	& ---			& 5.43  $\pm$ 0.18	& ---		  & ---  \\
 1032	& C     & 5.883 & 86.1  & Y     & 18.42 $\pm$ 0.17	& ---			& 4.94  $\pm$ 0.18	& ---		  & ---  \\
 1192	& A     & 5.643 & 195.1 & Y     & 12.52 $\pm$ 0.17	& ---			& 0.00  $\pm$ 0.18	& ---		  & ---  \\
 1192	& B     & 0.000 & 0.0   & Y     & 17.82 $\pm$ 0.17	& ---			& 5.30  $\pm$ 0.18	& ---		  & ---  \\
 1527	& A     & 0.000 & 0.0   & Y     & 14.73 $\pm$ 0.19	& ---			& 0.00  $\pm$ 0.18	& ---		  & ---  \\
 1527	& B     & 5.908 & 346.9 & Y     & 17.40 $\pm$ 0.19	& ---			& 2.67  $\pm$ 0.18	& ---		  & ---  \\
 1573	& A     & 0.000 & 0.0   & N     & 14.17 $\pm$ 0.17	& ---			& 0.00  $\pm$ 0.18	& ---		  & ---  \\
 1573	& B     & 3.933 & 300.1 & N     & 18.84 $\pm$ 0.17	& ---			& 4.68  $\pm$ 0.18	& ---		  & ---  \\
 1574	& A     & 0.000 & 0.0   & Y     & 14.26 $\pm$ 0.16	& ---			& 0.00  $\pm$ 0.18	& ---		  & ---  \\
 1574	& B     & 5.035 & 224.6 & Y     & 18.92 $\pm$ 0.17	& ---			& 4.66  $\pm$ 0.18	& ---		  & ---  \\            \hline
\end{longtable}
\tablefoot{Only objects with another comapnion at less than 3 arcsec have magnitudes in the SDSSz. KOI-0372 also has this information because during the observing run it seemed to have a close companion but after the reduction it was discarded.}
\end{landscape}
}


\begin{table*}
\caption{SED fitting results for the medium distance companions with 2MASS counterparts.}             
\setlength{\extrarowheight}{7pt}
\label{VOSAresults}      
\centering                          
\begin{tabular}{rcccccccrr}        
\hline\hline                 
KOI & Comp. &RA\tablefootmark{a} & DEC\tablefootmark{a}& $T_{eff}$  & log(g) & $\chi^2$ & SpT & Model \tablefootmark{b} & D(pc)\\
\hline                        
            131   &   A   &    299.097600  &      43.497601   &   6500.0    &     4.5    &    0.47     & F5  &   Kurucz  & $ 1120  ^{+      100  }_{-       120} $ \\
                  &   B   &    299.098259  &      43.496210   &   5500.0    &     4.5    &    0.90     & G8  &   Kurucz  & $ 2510  ^{+     330  }_{-      350} $ \\
                  &   B   &    299.098259  &      43.496210   &   5400.0    &     3.5    &    0.86     & G3  &  NextGen  & ---\tablefootmark{d} \\
            212   &   A   &    296.139750  &      41.603199   &   5800.0    &     4.5    &    0.92     & G4  &  NextGen  & $ 1010  ^{+    230  }_{-      130} $ \\
                  &   B   &    296.138708  &      41.604036   &   4400.0    &     4.5    &    5.66     & K5  &  NextGen  & $ 1760  ^{+    230  }_{-      240} $ \\
                  &   C   &    296.139026  &      41.601652   &   3400.0    &     4.5    &    0.61     & M4  &   COND00  & $  366  ^{+    136  }_{-      94} $ \\
                  &   B   &    296.138708  &      41.604036   &   4200.0    &     3.5    &    5.55     & K3  &  NextGen  & ---\tablefootmark{d} \\
                  &   C   &    296.139026  &      41.601652   &   3100.0    &     3.5    &    2.34     & $>$M6  &   COND00  & ---\tablefootmark{c} \\
            372   &   A   &    299.122500  &      41.866798   &   5750.0    &     4.5    &    1.21     & G5  &   Kurucz  & $  314  ^{+     85  }_{-      38} $ \\
                  &   B   &    299.123475  &      41.868137   &   4600.0    &     4.5    &    0.68     & K4  &  NextGen  & $ 1340  ^{+    230  }_{-      180} $ \\
                  &   B   &    299.123475  &      41.868137   &   4400.0    &     3.5    &    0.67     & K2  &  NextGen  & ---\tablefootmark{d} \\
            375   &   A   &    291.201150  &      51.144299   &   5800.0    &     4.5    &    1.60     & G4  &  NextGen  & $  495  ^{+    112  }_{-      65} $ \\
                  &   B   &    291.201750  &      51.142945   &   3500.0    &     4.5    &    1.41     & M3  &  NextGen  & $  831  ^{+    123  }_{-     212} $ \\
                  &   B   &    291.201750  &      51.142945   &   3100.0    &     3.5    &    3.04     & $>$M6  &   COND00  & ---\tablefootmark{c} \\
            520   &   A   &    294.667950  &      43.853298   &   5200.0    &     4.5    &    5.36     & K0  &  NextGen  & $  567  ^{+     66  }_{-      60} $ \\
                  &   B   &    294.666387  &      43.853345   &   3600.0    &     4.5    &   14.80     & M2  &  DUSTY00  & $ 1020  ^{+    170  }_{-     130} $ \\
                  &   B   &    294.666387  &      43.853345   &   3900.0    &     3.5    &   15.70     & K6  &   COND00  & ---\tablefootmark{d} \\
            623   &   A   &    295.226400  &      50.558998   &   6250.0    &     4.5    &    1.16     & F7  &   Kurucz  & $  411  ^{+     59  }_{-      60} $ \\
                  &   B   &    295.225822  &      50.557572   &   4800.0    &     4.5    &    2.41     & K3  &  NextGen  & $  439  ^{+     47  }_{-      55} $ \\
                  &   B   &    295.225822  &      50.557572   &   4400.0    &     3.5    &    1.88     & K2  &  NextGen  & $ 7920  ^{+    960  }_{-    1300} $ \\
            641   &   A   &    299.299500  &      40.235100   &   3600.0    &     4.5    &    1.29     & M2  &  DUSTY00  & $  117  ^{+     19  }_{-      15} $ \\
                  &   C   &    299.299054  &      40.234189   &   3400.0    &     4.5    &    1.60     & M4  &  NextGen  & $   76  ^{+     28  }_{-      20} $ \\
                  &   C   &    299.299054  &      40.234189   &   3700.0    &     3.5    &    1.68     & M1  &   COND00  & $10370  ^{+   1800  }_{-    2400} $  \\
            841   &   A   &    292.236750  &      41.085899   &   5400.0    &     4.5    &    1.17     & G9  &  NextGen  & $ 1170  ^{+    160  }_{-     150} $ \\
                  &   C   &    292.237780  &      41.087064   &   5000.0    &     4.5    &    1.53     & K1  &  NextGen  & $ 2120  ^{+   1070  }_{-     270} $ \\
                  &   C   &    292.237780  &      41.087064   &   4800.0    &     3.5    &    1.46     & G9  &  NextGen  & $10830  ^{+  1900   }_{-    1900}$ \\
           1527   &   A   &    296.671350  &      43.498402   &   5400.0    &     4.5    &    1.56     & G9  &  NextGen  & $  773  ^{+    103  }_{-      98} $ \\
                  &   B   &    296.670978  &      43.500000   &   4800.0    &     4.5    &    2.08     & K3  &  NextGen  & $ 1980  ^{+    220  }_{-     220} $ \\
                  &   B   &    296.670978  &      43.500000   &   4600.0    &     3.5    &    1.90     & K1  &  NextGen  & ---\tablefootmark{d} \\
\hline                                   
\end{tabular}
\tablefoot{ \\
\tablefoottext{a}{Calculated RA and DEC from the position of the KOI according to the distance and angle derived in this paper.}\\
\tablefoottext{b}{Best fit model used by VOSA. See http://svo.cab.inta-csic.es/theory/vosa/ for more information}\\
\tablefoottext{c}{Effective temperature is outside the model range}\\
\tablefoottext{d}{Estimated distance is greater than 21 Kpc (the estimated Milky Way limit in the Kepler line of sight)}
}
\end{table*}


\begin{table*}
\caption{New planet transit depth and relative radii to the parent star accounting for the blended objects in the Kepler images. In this calculation we have just taken into account the closest objects ($< 3$ arcsec) to the KOI.}             
\label{newparams}      
\centering                          
\begin{tabular}{c c c | c c c | c c }        
\hline\hline                 
         Planet ID   &    Cat. Depth   &   NewDepth    & Cat $R_p/R_*$	& New $R_p/R_*$\tablefootmark{a}  &  Sec. $R_p/R_*$\tablefootmark{b}  & Cat. $R_p$\tablefootmark{c}  & New $R_p$\tablefootmark{d}   \\    
         (KOI.Planet)	&	(ppm)	   &		(ppm)		&	 ($10^{-2}$)		&	 ($10^{-2}$)	&	 ($10^{-2}$) &	 $R_{Earth}$&	 $R_{Earth}$\\
\hline                        
         298.01 &      274 &   472$\pm$  67 &   1.41$\pm$0.04 &   2.2$\pm$0.2     &    2.6$\pm$  0.3 &      1.40 &     2.16    \\
         379.01 &      292 &   422$\pm$  24 &   1.6$\pm$0.1   &   2.1$\pm$0.1     &    3.1$\pm$  0.2 &      2.58 &     3.38    \\
         379.02 &      136 &   196$\pm$  11 &   1.1$\pm$0.1   &   1.40$\pm$0.04   &    2.1$\pm$  0.1 &      1.83 &     2.31    \\
         387.01 &     1122 &  1137$\pm$  75 &   3.3$\pm$0.3   &   3.4$\pm$0.1     &   29.4$\pm$ 73.4 &      2.18 &     2.23    \\
         401.01 &     2103 &  2363$\pm$ 132 &   4.1$\pm$0.2   &   4.9$\pm$0.1     &   13.8$\pm$  3.1 &      6.57 &     7.82    \\
         401.02 &     1618 &  1818$\pm$ 101 &   4.2$\pm$0.2   &   4.3$\pm$0.1     &   12.1$\pm$  2.7 &      6.67 &     6.85    \\
         433.01 &     2864 &  3048$\pm$  76 &   5.10$\pm$0.04 &   5.5$\pm$0.1     &   21.8$\pm$  4.2 &      5.60 &     6.06    \\
         433.02 &    13690 & 14570$\pm$ 365 &  11.7$\pm$0.1   &  12.1$\pm$0.2     &   47.6$\pm$  9.3 &     12.90 &    13.27    \\
         592.01 &      539 &   561$\pm$  19 &   2.6$\pm$0.1   &   2.37$\pm$0.04   &   11.7$\pm$  4.9 &      2.74 &     2.48    \\
         626.01 &      374 &   378$\pm$   4 &   1.8$\pm$0.1   &   1.94$\pm$0.01   &   19.2$\pm$  8.7 &      2.09 &     2.30    \\
         628.01 &      476 &   504$\pm$  21 &   2.2$\pm$0.2   &   2.24$\pm$0.05   &    9.3$\pm$  3.4 &      1.87 &     1.90    \\
         641.01 &     1002 &  2225$\pm$  90 &   3.1$\pm$1.3   &   4.7$\pm$0.1     &    4.3$\pm$  0.1 &      1.83 &     2.82    \\
         644.01 &    23950 & 30367$\pm$ 591 &  13.87$\pm$0.03 &  17.4$\pm$0.2     &   33.7$\pm$  1.2 &     33.16 &    41.67    \\
         645.01 &      201 &   239$\pm$   5 &   1.61$\pm$0.05 &   1.55$\pm$0.02   &    3.6$\pm$  0.2 &      2.53 &     2.44    \\
         645.02 &      257 &   305$\pm$   7 &   1.59$\pm$0.03 &   1.75$\pm$0.02   &    4.0$\pm$  0.2 &      2.51 &     2.75    \\
         658.01 &      505 &   517$\pm$   8 &   2.1$\pm$0.1   &   2.27$\pm$0.02   &   14.7$\pm$  5.0 &      2.03 &     2.18    \\
         658.02 &      484 &   496$\pm$   8 &   2.1$\pm$0.1   &   2.23$\pm$0.02   &   14.4$\pm$  4.9 &      2.02 &     2.13    \\
         658.03 &      166 &   170$\pm$   3 &   1.2$\pm$0.1   &   1.30$\pm$0.01   &    8.5$\pm$  2.9 &      1.14 &     1.25    \\
         703.01 &      130 &   131$\pm$   1 &   1.04$\pm$0.05 &   1.142$\pm$0.003 &   17.9$\pm$ 12.1 &      1.36 &     1.50    \\
         721.01 &      276 &   284$\pm$   2 &   1.63$\pm$0.03 &   1.685$\pm$0.007 &    9.9$\pm$  1.4 &      2.76 &     2.86    \\
         841.01 &     2967 &  3071$\pm$  45 &   5.4$\pm$0.1   &   5.54$\pm$0.04   &   29.6$\pm$  6.3 &      5.44 &     5.56    \\
         841.02 &     4962 &  5136$\pm$   7 &   7.0$\pm$0.2   &   7.2$\pm$0.1     &   38.3$\pm$  8.1 &      7.05 &     7.19    \\
        1375.01 &     2608 &  2651$\pm$  99 &   5.3$\pm$0.6   &   5.1$\pm$0.1     &   40.0$\pm$ 44.8 &      6.65 &     6.44    \\

\hline                                   
\end{tabular}

\tablefoot{KOI-0704.01 is not present due to the large errors in its $\Delta z$ value\\
\tablefoottext{a}{New planet-to-star radii ratio assuming no limb-darkening.}\\
\tablefoottext{b}{Planet to star radius assuming that the host is actually the secondary companion detected at less than 3 arcsec. }\\
\tablefoottext{c}{Planet radii calculated by the Kepler Team \citep{batalha12}.}\\
\tablefoottext{d}{Planet radii assuming the new depth and no limb-darkening. No error is presented since no error for the stellar radii is given.}
}
\end{table*}

\end{document}